\documentclass[11pt,a4paper]{article}
\pdfoutput=1

\usepackage{amsmath,amsfonts,mathrsfs,graphicx,bbm,bm}
\usepackage{fixltx2e}
\usepackage{cancel}

\usepackage[font=small]{caption}
\usepackage[text={450pt,650pt},headheight={14.5pt},centering]{geometry}

\usepackage{lmodern}

\usepackage{graphicx}
\usepackage{booktabs}
\usepackage{dsfont}

\usepackage{mathrsfs}
\usepackage{amsmath,amssymb}
\usepackage{mathtools}
\usepackage{bbm}
\usepackage{slashed}
\usepackage{braket}

\usepackage{comment}
\usepackage{hyperref}

\usepackage{xspace}
\hypersetup{
    colorlinks,
    linkcolor={red!0!black},
    citecolor={green!0!black},
    urlcolor={blue!0!black}
}

\usepackage{color}
\usepackage{tikz} 

\usetikzlibrary{plotmarks,calc,decorations,decorations.pathmorphing}
\usetikzlibrary{arrows}

\tikzstyle dynkin node=[very thick,shape=circle,draw,inner sep=0pt,minimum size=5mm]
\tikzstyle dynkin line=[very thick]
\tikzstyle inverse line=[gray,line width=1.46pt,line cap=round, dash pattern=on 0pt off 2\pgflinewidth]
\tikzstyle red phase=[red,decoration={snake,amplitude=0.1mm,segment length=1.6mm},decorate]
\tikzstyle blue phase=[blue,decoration={snake,amplitude=0.1mm,segment length=0.9mm},decorate]
\usetikzlibrary{calc}

\definecolor{grey}{rgb}{0.4,0.4,0.5}
\definecolor{darkgreen}{rgb}{0,0.5,0}
\definecolor{darkred}{rgb}{0.6,0.0,0}
\definecolor{lightbrown}{rgb}{1,0.9,0.8}
\definecolor{brown}{rgb}{0.6,0.3,0.3}
\definecolor{darkblue}{rgb}{0,0,0.8}
\definecolor{darkmagenta}{rgb}{0.5,0,0.5}

\numberwithin{equation}{section}

\makeatletter
 \let\old@startsection=\@startsection
 \let\oldl@section=\l@section
 \renewcommand{\@startsection}[6]{\old@startsection{#1}{#2}{#3}{#4}{#5}{#6\mathversion{bold}}}
 \renewcommand{\l@section}[2]{\oldl@section{\mathversion{bold}#1}{#2}}
\makeatother

\renewcommand{\geq}{\geqslant}

\def\XXint#1#2#3{{\setbox0=\hbox{$#1{#2#3}{\int}$}
    \vcenter{\hbox{$#2#3$}}\kern-.5\wd0}}

\newcommand{\alg}[1]{\mathfrak{#1}}

\def\be{\begin{equation}}
\def\ee{\end{equation}}

\newcommand{\bea}{\begin{equation}\begin{aligned}}
\newcommand{\eea}{\end{aligned}\end{equation}}
\newcommand{\bei}{\begin{itemize}}
\newcommand{\eei}{\end{itemize}}
\newcommand{\bee}{\begin{enumerate}}
\newcommand{\eee}{\end{enumerate}}
\newcommand{\bal}{\begin{equation}\begin{aligned}}
\newcommand{\eal}{\end{aligned}\end{equation}}

\newcommand{\su}{\alg{su}}

\newcommand{\gen}{\mathbf}

\renewcommand{\sin}{\text{sin}}
\renewcommand{\cos}{\text{cos}}

\newcommand{\dsbl}{[\![}
\newcommand{\dsbr}{]\!]}

\newcommand{\smL}{{\scriptscriptstyle{\text{L}}}}
\newcommand{\smR}{{\scriptscriptstyle{\text{R}}}}
\newcommand{\smI}{{\scriptscriptstyle{\text{I}}}}
\newcommand{\smJ}{{\scriptscriptstyle{\text{J}}}}

\newcommand{\hslashslash}{%
  \raisebox{.9ex}{%
    \scalebox{.7}{%
      \rotatebox[origin=c]{17}{$-$}%
    }%
  }%
}

\renewcommand{\k}{%
  {%
   \vphantom{k}%
   \ooalign{\kern-.05em\smash{\hslashslash}\hidewidth\cr$k$\cr}%
   \kern-.025em
  }%
}

\begin{document}

\thispagestyle{empty}

\begin{flushright}\footnotesize\ttfamily
DMUS-MP-17/10\\
NORDITA 2017-111
\end{flushright}
\vspace{5em}

\begin{center}
\textbf{\Large\mathversion{bold} 
$q$-Poincar\'e invariance of the $AdS_3/CFT_2$ $R$-matrix}

\vspace{2em}

\textrm{\large Riccardo Borsato${}^1$, Joakim Str\"omwall${}^{2}$ and Alessandro Torrielli${}^{2}$} 

\vspace{2em}

\begingroup\itshape
1. Nordita, Stockholm University and KTH Royal Institute of Technology,\\ Roslagstullsbacken 23, SE-106 91 Stockholm, Sweden\\[0.2cm]

2. Department of Mathematics, University of Surrey, Guildford, GU2 7XH, UK\par\endgroup

\vspace{1em}

\texttt{riccardo.borsato@su.se, j.stromvall@surrey.ac.uk, a.torrielli@surrey.ac.uk}


\end{center}

\vspace{6em}

\begin{abstract}\noindent
We consider the exact $R$-matrix of $AdS_3/CFT_2$, which is the building block for describing the scattering of worldsheet excitations of the light-cone gauge-fixed backgrounds $AdS_3\times S^3\times T^4$ and $AdS_3\times S^3\times S^3\times S^1$ with pure Ramond-Ramond fluxes.
We show that $R$ is invariant under a  ``deformed boost'' symmetry, for which we write an explicit exact coproduct, i.e. its action on 2-particle states.
When we include the boost, the symmetries of the $R$-matrix close into a $q$-Poincar\'e superalgebra. 
Our findings suggest that the recently discovered boost invariance in $AdS_5/CFT_4$ may be a common feature of $AdS/CFT$ systems that are treatable with the exact techniques of integrability.
With the aim of going towards a universal formulation of the underlying Hopf algebra, we also propose a universal form of the $AdS_3/CFT_2$ classical $r$-matrix.
\end{abstract}

\newpage

\tableofcontents


\section{\label{sec:level1}Introduction}

\subsection{Quantum group symmetries in $AdS/CFT$}

The progress in our understanding of the algebraic structure behind the AdS/CFT correspondence, and the integrability of its most symmetric incarnation \cite{Beisert:2010jr,Arutyunov:2009ga}, seems to be continuing as more examples are being systematically explored. The core of the method defines an  eigenvalue problem for the Hamiltonian of an effective two-dimensional integrable chain, and applies the Bethe ansatz to its exact $S$-matrix. Integrability is tied to a large algebra of non-abelian symmetries which form a Hopf superalgebra, and this makes it possible to ultimately solve the system via the tools of the representation theory of quantum groups.

The path to such a solution is however not a straightforward one, as these Hopf superalgebras are rather exotic. They display an infinite tower of generators labelled by an integer, and are very close to Yangian algebras \cite{Drinfeld:1985rx,Drinfeld:1986in,Molev:1994rs,Khoroshkin:1994uk,Molev:2003,MacKay:2004tc,Beisert:2007ds,Torrielli:2011gg}. The level 0 typically coincides with the manifest superconformal symmetry of the theory, partially broken and centrally-extended {\it \`a la} Beisert \cite{Beisert:2005tm,Arutyunov:2006ak}. The central extension goes hand-in-hand with certain non-linear constraints on the central charges, which, in turn, are linked to deformations appearing in the Hopf-algebra coproduct map \cite{Gomez:2006va,Plefka:2006ze}. Furthermore, the Yangian \cite{Beisert:2007ds} displays extra generators \cite{Matsumoto:2007rh,Beisert:2007ty} with no level-0 analog. These symmetries have been dubbed {\it secret} or {\it bonus}. They have also been observed in boundary scattering problems \cite{Regelskis:2011fa}, $n$-point amplitudes \cite{Beisert:2011pn}, the pure-spinor formalism \cite{Berkovits:2011kn}, in the quantum-affine deformations \cite{deLeeuw:2011fr} and in the context of Wilson loops \cite{Munkler:2015xqa}. This makes it quite a significant feature of the system and not an isolated instance \cite{deLeeuw:2012jf}.

Some light on the problem was recently shed by applying the so-called $R\mathcal T\mathcal T$ formulation \cite{Beisert:2014hya}. In this approach, one starts from  the $S$-matrix in the fundamental representation, and generates from it an algebra of symmetries of the integrable system at hand. In the process, the operator deforming the coproduct was re-interpreted as a particular Yangian generator of level $-1$. This has a correspondent in the classical $r$-matrix algebra \cite{Beisert:2007ty} constructed in the spirit of Drinfeld's second realisation of Yangians \cite{Drinfeld:1987sy,Spill:2008tp}. 

Even with this step, the accommodation of the full quantum-group tower of symmetries appears still out of reach, and the hope to find the universal $R$-matrix and have a full control of the representation theory \cite{Beisert:2006qh,Arutyunov:2009pw}, relies on further progress. It has very recently become clear in fact \cite{Beisert:2016qei,Beisert:2017xqx} that extra generators (automorphisms) cannot be done without. In \cite{Borsato:2017lpf} an entirely different generator was found for superstrings in $AdS_5 \times S^5$, as we will describe in a subsection below.  This is the $5D$ case, with a dual theory given by $4D$ ${\cal{N}}=4$ super Yang-Mills.

Analogous non-standard quantum algebras and associated bonus generators have been found in lower-dimensional $AdS/CFT$ as well. All these settings share peculiar algebraic features stemming from the vanishing of the Killing form of their superisometry \cite{Zarembo:2010sg,Wulff:2015mwa}, dictated by string coset integrability and $\sigma$-model scale invariance. This seems to tie in with the algebraic peculiarities we have been discussing, which, albeit with a richness of variants, appear to carry over to all cases.
From a quantum-group viewpoint, the integrable structure behind the $AdS_4$ {case \cite{Arutyunov:2008if,Stefanski:2008ik,Klose:2010ki} is} reduced for the most part to the five-dimensional case (although the physics is very different).

The $AdS_3/CFT_2$ integrability \cite{Babichenko:2009dk,Sfondrini:2014via,Borsato:2016hud}---see also \cite{Pesando:1998wm,OhlssonSax:2011ms,Sundin:2012gc}---provides another fertile realisation of these exotic group-theory structures \cite{Regelskis:2015xxa}. This is the setup in which we will work in this paper.
The  program {of integrability is } carried on for superstrings on $AdS_3\times S^3\times S^3\times S^1$ and on $AdS_3\times S^3\times T^4$.
The former background contains a parameter $\alpha$ corresponding to the relative radii of the $S^3$s, reflected in the superisometry algebra $\alg{d}(2,1;\alpha)_\smL\oplus \alg{d}(2,1;\alpha)_\smR$. Here L and R label the two copies. An $\alpha \to 0$ contraction produces $\alg{psu}(1,1|2)_\smL\oplus\alg{psu}(1,1|2)_\smR$, the superisometry algebra of the latter background. The bonus symmetry was found in \cite{Pittelli:2014ria}, cf. \cite{Regelskis:2015xxa}. 
{Before discussing the results in $AdS_3/CFT_2$ in more details, let us review the boost invariance that was identified in $AdS_5/CFT_4$.}

\subsection{Deformed Poincar\'e supersymmetry in $AdS_5/CFT_4$}

In \cite{Borsato:2017lpf}, a new symmetry of the $AdS_5/CFT_4$ $S$-matrix was found, realising the boost of a specific $q$-deformation of $1+1$-dimensional Poincar\'e superalgebra. Other $q$-deformations have appeared in \cite{Hoare:2011fj,Beisert:2008tw,Klimcik:2008eq,Delduc:2013qra,Sfetsos:2013wia,Hollowood:2014qma,Pachol:2015mfa,
Arutynov:2014ota,Borsato:2016hud}. In these parallel lines of investigation, however, the $q$-deformation is {\it superimposed} to the algebra, and deforms the theory. This is not what we study in this context, where the super $q$-Poincar\'e deformation is part of the ordinary superstring theory. Boost operators on spin-chains have a long history \cite{tetelman1982lorentz,sogo1983boost,Bargheer:2008jt}. See also \cite{Beisert:2007jv,Zwiebel:2008gr} in the context of long-range spin-chains, \cite{Kawaguchi:2011wt,Kawaguchi:2012ug,Kawaguchi:2013lba} in the study of sigma models, and  \cite{Beisert:2016qei,Klose:2016uur,Klose:2016qfv} in the development of algebraic methods for AdS/CFT integrability.  

The paper \cite{Gomez:2007zr} was the first to investigate remnants of the Poincar\'e algebra  in the $AdS_5/CFT_4$ integrable problem. The exact dispersion relation of the excitations  was interpreted as the Casimir of a $q$-Poincar\'e algebra:
\begin{equation}
C=\gen{H}^2+g^2 (\gen{K}^{\frac{1}{2}}-\gen{K}^{-\frac{1}{2}})^2, 
\end{equation}
where $\gen H$ is the generator corresponding to the energy and $\gen K=\exp(i\gen P)$ is the exponential of the worldsheet momentum. The coupling $g$, which is the tension of the string, plays the role of the deformation parameter of $q$-Poincar\'e.
The {\it boost} generator $\gen{J}$ was introduced as producing shifts 
\begin{equation}
\gen{J}: z \to z + c
\end{equation}
in the torus variable $z$ that uniformises the dispersion relation~\cite{Janik:2006dc}.
Immediately afterwards, the paper \cite{Young:2007wd} generalised this construction to the full centrally-extended $\alg{psu}(2|2)$ algebra, under which the $AdS_5/CFT_4$ $S$-matrix is invariant. Nevertheless, a coproduct  was given only for a carefully selected subalgebra of generators, and it turned out to be incompatible with the $S$-matrix---e.g. the energy was not co-commutative. 

In \cite{Borsato:2017lpf}, it was demonstrated that one can overcome these shortcomings by allowing a non-standard coproduct for $\gen J$, in such a way that the boost is a symmetry of the $S$-matrix, as well as all other generators in the superalgebra. In section \ref{sec:algebra} we adopt this strategy to extend these results to $AdS_3/CFT_2$.

\subsection{Deformed Poincar\'e supersymmetry in $AdS_3/CFT_2$}

The global superconformal symmetries of the  $AdS_3 \times S^3 \times S^3 \times S^1$ and $AdS_3 \times S^3 \times T^4$ backgrounds are broken by a choice of vacuum. This corresponds to fixing light-cone gauge compatibly with the BMN ground state, or in the spin-chain picture to the choice of the reference state. The elementary excitations above the vacuum transform in the little group of residual symmetry which preserves the vacuum. These residual symmetries consist of $2$ copies of the centrally extended $\alg{su}(1|1)$ superalgebra in the case of $AdS_3 \times S^3 \times S^3 \times S^1$~\cite{Borsato:2012ud,Borsato:2012ss,Borsato:2015mma}, while $4$ copies in the case of $AdS_3 \times S^3 \times T^4$~\cite{Borsato:2013qpa,Borsato:2014exa,Borsato:2014hja,Lloyd:2014bsa}{---see also \cite{Rughoonauth:2012qd,Abbott:2012dd,Beccaria:2012kb,Sundin:2013ypa,Abbott:2014rca,Abbott:2015pps,Prinsloo:2015apa,Abbott:2015mla,Sundin:2016gqe}, and} section~\ref{sec:algebra} for more details. 
The $AdS_3/CFT_2$ integrable problem contains not only massive but also massless excitations. This appears as a novel feature compared to the $AdS_5/CFT_4$ case, and it offers both challenges {(such as mismatches with perturbation theory waiting to be fully resolved \cite{Sundin:2016gqe})} and interesting physics~\cite{Sax:2012jv}, {see also \cite{Eberhardt:2017fsi,Baggio:2017kza,Eberhardt:2017pty}.}

By adopting the spirit of \cite{Gomez:2007zr,Young:2007wd}, in \cite{Stromwall:2016dyw} it was shown that for massless excitations of the above $AdS_3$ backgrounds the corresponding residual symmetries can be extended to a $q$-Poincar\'e superalgebra analogous to that of $AdS_5/CFT_4$. Due to the massless dispersion relation, the $q$-deformed energy coproduct turns out to be co-commutative, hence an exact symmetry of the $S$-matrix. This new interpretation of the  magnon supersymmetry in the massless case also allows a very concise reformulation of the comultiplication map, and connections with the scattering of phonons \cite{Ballesteros:1999ew}. Matching more closely with the relativistic theory might be of significance to describe certain limits of the putative dual field theories \cite{Sax:2014mea,Tong:2014yna}.
In this setting, however, the boost-coproduct is not a symmetry of the $S$-matrix, but it rather annihilates it. 
This was showed in \cite{Fontanella:2016opq}, where an associated differential-geometric framework was then proposed based on a flat would-be connection. 
It seems that the construction of \cite{Stromwall:2016dyw,Fontanella:2016opq} is limited to the case of massless excitations, and it is not clear how to extend it to representations of generic mass. 

\vspace{12pt}

In this paper we adopt a point of view close to~\cite{Borsato:2017lpf}, and our discussion of the $q$-Poincar\'e supersymetry is valid for generic values of the mass. 
In section~\ref{sec:algebra} we construct a coproduct for the boost, and  we check that it is a symmetry of the $S$-matrix in the relevant $2$-particle representations.
We also study how the boost transforms under crossing transformations.
In section~\ref{sec:sem-lim} we take the semiclassical limit of the deformed superalgebra, which yields a classical Lie superalgebra that may be obtained also as a contraction of $\alg{sl}(1|2)$.
In section~\ref{sec:univ-r} we write down a proposal for a universal classical $r$-matrix that matches with the known results in the fundamental representation, and we check that it satisfies the classical Yang-Baxter equation in universal form.
We use this result in section~\ref{sec:cobra} to compute the cobrackets of the generators, including the boost.

\section{Symmetry algebra and the boost}\label{sec:algebra}
In this section we will review in more detail the superalgebra of symmetries of the $AdS_3/CFT_2$ integrable models, and its short fundamental representations. The formulation we employ here differs from that of~\cite{Borsato:2012ud,Borsato:2013qpa} in how we treat the off-shell central extension; in particular, instead of introducing the central elements $\gen C,\overline{\gen C}$, we write the results of the corresponding anticommutators just in terms of the momentum generator $\gen P$, or more conveniently in terms of\footnote{To avoid confusion coming from different notations used in the literature, we stress that  $\gen P,\gen K$ are respectively the worldsheet momentum and its exponential (multiplied by $i$).}  $\gen{K}\equiv\exp(i\gen{P})$.
As done in~\cite{Young:2007wd,Borsato:2017lpf} for the $AdS_5/CFT_4$ case, we therefore formulate the symmetry algebra as a deformation of the universal enveloping algebra. In this formulation $g$, which at large values of the tension of the string is the tension itself, plays the role of the deformation parameter. 

As recalled in the introduction, after fixing light-cone gauge on the worldsheet for the  $AdS_3\times S^3\times S^3\times S^1$ background one ends up with a centrally extended  $\su(1|1)_\smL\oplus \su(1|1)_\smR$ superalgebra, where the labels Left (L) and Right (R) distinguish the two copies. Worldsheet excitations are organised in 2-dimensional irreducible representations of this superalgebra. They carry labels L or R (see below) which remind us that on-shell ($\gen P=0$) only the L (resp. R) copy of the superalgebra acts non-trivially on L (resp. R) excitations. Their masses can take only the  values $m=0,\alpha,1-\alpha,1$.
The construction carried on for the $AdS_3\times S^3\times T^4$ background leaves instead a larger symmetry algebra. This may be obtained by considering two copies of the above centrally extended $\su(1|1)_\smL\oplus \su(1|1)_\smR$ superalgebra, where we mod out half of the central elements to leave only their symmetric combinations, and the odd generators are organised in (anti)fundamental representations of an additional $\su(2)$ symmetry.
The worldsheet excitations are still labelled by L and R, and their masses can be just $m=0,1$.
To keep the discussion as general as possible, in the following we will consider just one copy of the centrally extended $\su(1|1)_\smL\oplus \su(1|1)_\smR$ superalgebra, and we will consider L and R representations of generic mass $m$. Therefore, in order to obtain the results for the $AdS_3\times S^3\times S^3\times S^1$ background it will be enough to set the masses to the desired values. The results for the $AdS_3\times S^3\times T^4$ background are instead obtained by constructing the bifundamental representations as explained in~\cite{Borsato:2013qpa,Borsato:2014hja}, see in particular section 3.1 of~\cite{Borsato:2014hja}.

The centrally extended  $\su(1|1)_\smL\oplus \su(1|1)_\smR$ superalgebra is spanned by the supercharges $\gen Q_\smI,\overline{\gen Q}_\smI$ and the central elements $\gen H_\smI, \gen P$ (here the subscript I$=$L,R denotes the two copies), which close into the anticommutation relations
\be\label{eq:comm-rel-noJ-AdS3}
\begin{aligned}
&\{\gen{Q}_\smL,\overline{\gen Q}_\smL\}=\gen{H}_\smL,
\qquad &&\{\gen{Q}_\smL,\gen{Q}_\smR\}=\tfrac{ig}{2}\left(\gen{K}^{\frac{1}{2}}-\gen{K}^{-\frac{1}{2}}\right),\\
&\{\gen{Q}_\smR,\overline{\gen Q}_\smR\}=\gen{H}_\smR,
\qquad &&\{\overline{\gen{Q}}_\smL,\overline{\gen{Q}}_\smR\}=\tfrac{ig}{2}\left(\gen{K}^{\frac{1}{2}}-\gen{K}^{-\frac{1}{2}}\right).
\end{aligned}
\ee
Useful combinations are the Hamiltonian $\gen{H}=\gen{H}_\smL+\gen{H}_\smR$, and the central charge $\gen{M}=\gen{H}_\smL-\gen{H}_\smR$ which, as we will recall later, is related to the mass. The two copies of $\su(1|1)$ decouple on-shell, i.e. when $\gen P= 0$.

We now introduce a boost generator $\gen J$ such that
\be\label{eq:comm-rel-J-AdS3}
\begin{aligned}
& [\gen{J},\gen{P}]=i\, \gen{H},\qquad\qquad
&&[\gen{J},\gen{Q}_\smI]=-\tfrac{ig}{4}\left(\gen{K}^{\frac{1}{2}}+\gen{K}^{-\frac{1}{2}}\right) \overline{\gen{Q}}_{\bar{\smI}},\\
&[\gen{J},\gen{H}]=\tfrac{g^2}{2}\left(\gen{K}-\gen{K}^{-1}\right) ,\qquad
&&[\gen{J},\overline{\gen{Q}}_\smI]=-\tfrac{ig}{4}\left(\gen{K}^{\frac{1}{2}}+\gen{K}^{-\frac{1}{2}}\right) \gen{Q}_{\bar{\smI}},
\end{aligned}
\ee
where I$=$L,R and $\bar{\text{L}}=\text{R},\bar{\text{R}}=\text{L}$. A difference with respect to the construction of~\cite{Stromwall:2016dyw,Fontanella:2016opq} is that here we do not introduce a boost generator for each copy L and R, we rather have one common boost relating the two copies. The boost also breaks the centrality of $\gen H$ and $\gen P$. The above commutation relations are also compatible with the automorphism $\gen b$ acting only on the supercharges as
\be
[\gen{b},\gen{Q}_\smL]=-2\gen{Q}_\smL,\quad
[\gen{b},\overline{\gen Q}_\smL]=+2\overline{\gen Q}_\smL,\qquad
[\gen{b},\gen{Q}_\smR]=+2\gen{Q}_\smR,\quad
[\gen{b},\overline{\gen Q}_\smR]=-2\overline{\gen Q}_\smR.
\ee
The generator $\gen b$ is the only combination of the two $\gen b_\smI$ outer automorphisms of $\su(1|1)_\smI$ that survive after introducing the central extension ($\gen P\neq 0$). For convenience, we summarise our conventions for the generators we shall use and their fermionic degree in the following table:

\begin{center}
\label{tabella}
 \begin{tabular}{||c c||} 
 \hline
 Gener. & Degree  \\ [0.5ex] 
 \hline\hline
 $\gen{Q}_\smL$ & 1  \\ 
 \hline
 $\overline{\gen{Q}}_\smL$ & 1  \\
 \hline
 $\gen{Q}_\smR$ & 1  \\
 \hline
 $\overline{\gen{Q}}_\smR$ & 1  \\
 \hline
 $\gen{P}$ & 0  \\  
 \hline
 $\gen{K}$ & 0  \\ 
 \hline 
$\gen{H}$ & 0  \\ 
 \hline
$\gen{M}$ & 0  \\
 \hline 
$\gen{J}$ & 0  \\
 \hline
$\gen{b}$ & 0  \\ 
 \hline 
$\hat{\gen{B}}$ & 0  \\ 
 \hline
 \end{tabular}
\end{center} 
The generator $\hat{\gen{B}}$ will appear later --- cf. (\ref{eq:secret}). The Casimir of the $q$-Poincar\'e subalgebra (generated by $\gen{H},\gen{P},\gen{J}$) is denoted by $C=\gen{H}^2+g^2 (\gen{K}^{\frac{1}{2}}-\gen{K}^{-\frac{1}{2}})^2$. When comparing it to the shortening condition $\gen{H}^2=\gen M^2-g^2 (\gen{K}^{\frac{1}{2}}-\gen{K}^{-\frac{1}{2}})^2$ given in~\cite{Borsato:2012ud,Borsato:2013qpa} we see that we should set $C=\gen M^2$.

The short irreducible representations of the centrally extended $\su(1|1)_\smL\oplus \su(1|1)_\smR$ are 2-dimensional. They are labelled by three parameters\footnote{In the spirit of the original paper~\cite{Borsato:2012ud} we prefer to denote by $m$ the mass of the excitations, so that $m\geq 0$. In~\cite{Borsato:2015mma} $m$ was used for the eigenvalue of $\gen M$ (at $q=0$), which is positive/negative on L/R representations; in that case the mass of the excitations would be $|m|$.}: the mass $m$, the momentum $p$ and the coupling $g$. We will be interested in the L and the R representations\footnote{The two additional representations denoted by $\widetilde{\varrho}_\smL,\widetilde{\varrho}_\smR$ in~\cite{Borsato:2014hja} are simply obtained from the above ones by exchanging the roles of the bosons and the fermions.} $\varrho_\smL$ and $\varrho_\smR$, each spanned by a boson $\phi^\smI$ and a fermion $\psi^\smI$. On the (reducible) representation $\varrho_\smL\oplus\varrho_\smR=\text{span}\{\phi^\smL,\psi^\smL,\phi^\smR,\psi^\smR\}$ the above generators may be realised as explicit $4\times 4$ matrices
\be\label{eq:repr-Q}
\begin{aligned}
&\gen Q_\smL=  a_p\sigma_-\oplus b_p\, \sigma_+,
\qquad
&&\overline{\gen Q}_\smL=  \bar a_p\sigma_+\oplus \bar b_p\, \sigma_-,
\\
&\gen Q_\smR=  b_p\, \sigma_+\oplus a_p\sigma_-,
\qquad
&&\overline{\gen Q}_\smR= \bar b_p\, \sigma_- \oplus \bar a_p\sigma_+,
\end{aligned}
\ee
where $\sigma_\pm=\frac{1}{2}(\sigma_1\pm i \sigma_2)$. The L and R representations may be mapped to each other by swapping the labels L$\leftrightarrow$R on the charges and on the states. We take
\be
a_p=\bar a_p=\sqrt{\frac{g}{2}}\gamma_p,\qquad
b_p=\bar b_p=i\sqrt{\frac{g}{2}} \gamma_p^{-1}\left(\left(\tfrac{x^+}{x^-}\right)^{1/2}-\left(\tfrac{x^+}{x^-}\right)^{-1/2} \right),\qquad
\gamma_p=\sqrt{i(x^-_p-x^+_p)},
\ee
 and we make use of the Zhukovski variables $x^\pm_p$ which satisfy 
\be
x^+_p+\frac{1}{x^+_p}-x^-_p-\frac{1}{x^-_p}=\frac{2im}{g},\qquad\quad
\frac{x^+_p}{x^-_p}=e^{ip}.
\ee
Notice the dependence on the mass $m$ in the first of the above constraints.
One also finds
\be
\gen H = h_p \left[\gen 1_2\oplus \gen 1_2\right],\qquad
\gen M= m \left[\gen 1_2\oplus \left(- \gen 1_2\right)\right],\qquad
\gen b=  \sigma_3\oplus \left(- \sigma_3\right),
\ee
with
\be
h_p=\frac{ig}{2}\left(x^-_p-x^+_p+\frac{1}{x^+_p}-\frac{1}{x^-_p}\right)=\sqrt{m^2+4g^2\sin^2\frac{p}{2}}.
\ee
The sign of the eigenvalue of $\gen M$ allows us to distinguish between the L and R representations. The action of the generator $\gen b$ also differs on the two representations by a sign.
Finally, the boost is realised as $\gen J=i\gen H\partial_p$.

\vspace{12pt}

In~\cite{Borsato:2012ud} an $R$-matrix in the fundamental representation was found by demanding that it should be invariant under the symmetries, with the exception of the boost --- the reader is also referred to \cite{Dabrowski:1991nr,Hinrichsen:1991nj,Bracken:1994hz}. In our conventions, when scattering the tensor-product representation $\varrho\otimes\chi$, the symmetry invariance of the $R$-matrix is imposed as
\be\label{eq:inv-R}
\Delta_{\chi\otimes\varrho}^{op}(\gen q)R = R\Delta_{\varrho\otimes\chi}(\gen q),
\ee 
where we use  the subscript to specify the tensor-product representation on which we should evaluate the coproduct, and we define $\Delta_{\chi\otimes\varrho}^{op} \equiv \Pi_g \Delta_{\chi\otimes\varrho} \Pi_g$ with $\Pi_g$ the graded permutation\footnote{In an explicit matrix realisation, when defining the ``${op}$'' of a coproduct one should also take care of swapping the labels $\{p_1,m_1\}\leftrightarrow\{p_2,m_2\}$ everywhere. With these conventions, the states are ordered as $\{(p_1,m_1),(p_2,m_2)\}$ both before and after the action of the $R$-matrix.}.
The coproduct that we use here is the one in the \emph{most symmetric frame}
\be\label{eq:Delta-su(1|1)2}
\begin{aligned}
&\Delta(\gen{Q}_\smI)=\gen{Q}_\smI\otimes \gen K^{-\frac{1}{4}}+\gen K^{\frac{1}{4}}\otimes \gen{Q}_\smI,\qquad
&&\Delta(\gen{H})=\gen{H}\otimes \gen 1+\gen 1\otimes \gen{H},\\
&\Delta(\overline{\gen Q}_\smI)=\overline{\gen Q}_\smI\otimes \gen K^{\frac{1}{4}}+\gen K^{-\frac{1}{4}}\otimes \overline{\gen Q}_\smI,\qquad
&&\Delta(\gen{M})=\gen{M}\otimes \gen 1+\gen 1\otimes \gen{M},\\
&\Delta(\gen{b})=\gen{b}\otimes \gen 1+\gen 1\otimes \gen{b},
&&\Delta(\gen P)=\gen{P}\otimes \gen 1+\gen 1\otimes \gen{P}.
\end{aligned}
\ee
The $R$-matrix is decomposed into blocks related by LR-symmetry, see~\cite{Borsato:2012ud,Borsato:2013qpa}. The two independent blocks are LL and LR, and one finds
\be\label{eq:R-LL}
\begin{aligned}
&R\ket{\phi^\smL \phi^\smL} = \ket{\phi^\smL \phi^\smL}, \qquad &&
R\ket{\phi^\smL \psi^\smL} = a^{\smL\smL}_{12}\ket{\phi^\smL \psi^\smL}+b^{\smL\smL}_{12}\ket{\psi^\smL \phi^\smL}, \\
&R\ket{\psi^\smL \psi^\smL} = c^{\smL\smL}_{12}\ket{\psi^\smL \psi^\smL}, &&
R\ket{\psi^\smL \phi^\smL} = (a^{\smL\smL}_{21})^*\ket{\psi^\smL \phi^\smL}+(b^{\smL\smL}_{21})^*\ket{\phi^\smL \psi^\smL}, 
\end{aligned}
\ee
\be\label{eq:R-LR}
\begin{aligned}
&R\ket{\phi^\smL \phi^\smR} = a^{\smL\smR}_{12}\ket{\phi^\smL \phi^\smR}+ b^{\smL\smR}_{12}\ket{\psi^\smL \psi^\smR}, \qquad &&
R\ket{\phi^\smL \psi^\smR} = \ket{\phi^\smL \psi^\smR}, \\
&R\ket{\psi^\smL \psi^\smR} = a^{\smL\smR}_{21}\ket{\psi^\smL \psi^\smR}+ b^{\smL\smR}_{21}\ket{\phi^\smL \phi^\smR}, &&
R\ket{\psi^\smL \phi^\smR} = c^{\smL\smR}_{12}\ket{\psi^\smL \phi^\smR}, 
\end{aligned}
\ee
where $*$ denotes complex conjugation---under which $(x^\pm)^*=x^\mp$. Here we have chosen an arbitrary normalisation by setting one element in each block to 1; the remaining coefficients are
\be
\begin{aligned}
&a^{\smL\smL}_{12}=\left(\frac{x^+_1}{x^-_1}\right)^{-1/2}\frac{ x^+_2-x^+_1}{x^+_2-x^-_1},\qquad
&&b^{\smL\smL}_{12}=\left({\frac{x^+_1}{x^-_1}}\right)^{-1/4}
\left({\frac{x^+_2}{x^-_2}}\right)^{1/4}\frac{ x^+_2-x^-_2}{x^+_2-x^-_1}\ \frac{\gamma_1 }{\gamma_2},
\\
&c^{\smL\smL}_{12}=\left({\frac{x^+_1}{x^-_1}}\right)^{-1/2}
   \left({\frac{x^+_2}{x^-_2}}\right)^{1/2}\frac{x^-_2-x^+_1}{x^+_2-x^-_1},
\end{aligned}
\ee
and
\be
\begin{aligned}
&a^{\smL\smR}_{12}=\left(\frac{x^+_1}{x^-_1}\right)^{-1/2} \frac{ x^-_2x^+_1-1}{x^-_1 x^-_2-1},
\qquad
b^{\smL\smR}_{12}=\left(\frac{x^+_2}{x^-_2}\right)^{-1/4}\left(\frac{x^+_1}{x^-_1}\right)^{-1/4}\frac{i \gamma_1 \gamma_2}{ x^-_1x^-_2-1},\\
&c^{\smL\smR}_{12}=\left(\frac{x^+_1}{x^-_1}\right)^{-1/2} \left(\frac{x^+_2}{x^-_2}\right)^{-1/2} 
\frac{ x^+_1x^+_2-1}{x^-_1 x^-_2-1}.
\end{aligned}
\ee
Braiding unitarity is written as $R^{op}R=1$, and one may check that  the Yang-Baxter equation is satisfied; a convenient way to check it is done by introducing the $S$-matrix $S=\Pi_g R$ so that
\be
S_{12}(p_2,p_3)S_{23}(p_1,p_3)S_{12}(p_1,p_2)=S_{23}(p_1,p_2)S_{12}(p_1,p_3)S_{23}(p_2,p_3).
\ee
The subscripts denote the subspaces on which the $S$-matrix is acting, e.g. $S_{12}=S\otimes \gen 1$, and one should take care of evaluating the $S$-matrix in the relevant representation.

\vspace{12pt}

As discovered in~\cite{Pittelli:2014ria}, one may identify a secret symmetry similar to the one appearing in the case of $AdS_5/CFT_4$.
The antisymmetric combination of the L and R secret symmetries of~\cite{Pittelli:2014ria} (at level 0) should be identified with our automorphism $\gen b$. The symmetric combination instead may be identified with $\hat{\gen B}$, the counterpart of the $AdS_5/CFT_4$ secret symmetry. 
{See also~\eqref{eq:dictionary-Antonio} for the explicit relation to generators used in the literature.}
In the $\varrho_\smL\oplus\varrho_\smR$ fundamental representation we write $\hat{\gen B}$ as
\be\label{eq:secret}
\hat{\gen{B}}=\frac{1}{4}\left(x^+_p+x^-_p-\frac{1}{x^+_p}-\frac{1}{x^-_p}\right)\ (\sigma_3\oplus\sigma_3),
\ee
which is compatible with the commutation relations
\be
[\hat{\gen B},\gen Q_\smI]= -\hat{\gen Q}_{\smI}-\left(\gen K^{\frac{1}{2}}+\gen K^{-\frac{1}{2}}\right)\ \overline{\gen{Q}}_{\bar{\smI}},
\qquad\qquad
[\hat{\gen B},\overline{\gen Q}_\smI]= \hat{\overline{\gen Q}}_{\smI}+\left(\gen K^{\frac{1}{2}}+\gen K^{-\frac{1}{2}}\right)\ \gen{Q}_{\bar{\smI}}.
\ee
Here hatted supercharges denote the ones at level 1 of the Yangian. We assume that we can use evaluation representation and identify e.g. $\hat{\gen Q}_{\smI}\sim \hat u \gen Q_{\smI}$ with $\hat u=(x^++x^-+1/x^++1/x^-)/2$.
One may check that the coproduct 
\be
\Delta(\hat{\gen{B}})=\hat{\gen{B}}\otimes \gen 1+\gen 1\otimes \hat{\gen{B}}+ \frac{i}{g}\sum_{\smI=\smL,\smR}\left( \gen K^{-\frac{1}{4}}\gen Q_{\smI}\otimes \gen K^{-\frac{1}{4}}\overline{\gen Q}_{\smI} 
+\gen K^{\frac{1}{4}}\overline{\gen Q}_{\smI}\otimes  \gen K^{\frac{1}{4}}\gen Q_{\smI} \right)
\ee
gives a symmetry of the $R$-matrix, both in the LL and the LR representations.

\vspace{12pt}

In order to determine the coproduct for the boost we follow the strategy used  in~\cite{Borsato:2017lpf} in the case of $AdS_5/CFT_4$: we constrain an appropriate Ansatz for the boost coproduct by imposing commutation relations~\eqref{eq:comm-rel-J-AdS3}, while using the above coproducts for all other generators in the algebra. 
The coproduct that we find \emph{in the fundamental representation}
\begin{equation}\label{eq:start-Delta-boost}
\Delta(\gen{J}) = \Delta'(\gen{J})+\mathcal{T}
\end{equation}
has obvious analogies with the one of~\cite{Borsato:2017lpf}.
In particular, the contribution $\Delta'(\gen{J})$ remains the same, since it is found by imposing commutation relations of the bosonic $q$-Poincar\'e subalgebra. One has
\be\label{eq:Deltap}
\begin{aligned}
&\Delta'(\gen{J})=\left(1-\frac{ {s}_{12}}{h_1}\right)\gen{J}\otimes \gen 1+\left(1+\frac{{s}_{12}}{h_2}\right)\gen 1\otimes \gen{J},\qquad
{s}_{12}=\frac{g}{2}\frac{\sin p_1+\sin p_2-\sin (p_1+p_2)}{w_1^{-1}-w_2^{-1}},
\end{aligned}
\ee
where
\be
\begin{aligned}
w_p&=\frac{2\, h_p}{g\ \sin p}=2\frac{1+x^-_px^+_p}{x^-_p+x^+_p}.
\end{aligned}
\ee
The tail $\mathcal T$ is obtained by imposing commutation relations between $\gen J$ and the supercharges, and we find
\begin{eqnarray}
&&\mathcal{T}=\mathcal{T}_{\gen{H}\hat{\gen{B}}}+\mathcal{T}_{\gen{M}\gen{b}}+\mathcal{T}_{\smL}+\mathcal{T}_{\smR}+\mathcal{T}_1,\label{eq:DeltaJ}\\ 
&& \nonumber \\
&&\mathcal{T}_{\gen{H}\hat{\gen{B}}}=\frac{1}{2} \frac{1}{w_1-w_2}\left(1-\tan\frac{p}{2}\otimes \tan\frac{p}{2}\right) \left(\gen{H}\otimes \hat{\gen{B}}+ \hat{\gen{B}}\otimes \gen{H}\right),\nonumber \\ 
&&\mathcal{T}_{\gen{M}\gen{b}}=\frac{1}{8}\frac{w_1+w_2}{w_1-w_2}\left( \gen{M}\otimes \gen{b}+\gen{b}\otimes\gen{M} \right)\nonumber\\
&&\mathcal{T}_{\smJ}=\frac{1}{2}\frac{w_1+w_2}{w_1-w_2}\left(  \gen K^{-\frac{1}{4}}\gen{Q}_{\smJ}\otimes  \gen K^{-\frac{1}{4}} \overline{\gen{Q}}_{\smJ}- \gen K^{\frac{1}{4}} \overline{\gen{Q}}_{\smJ}\otimes  \gen K^{\frac{1}{4}}\gen{Q}_{\smJ} \right)\nonumber
\end{eqnarray}
Notice the strong analogies with the $AdS_5/CFT_4$ result of~\cite{Borsato:2017lpf} when looking at the bilinear piece in supercharges and the contribution  with the secret symmetry $\hat{\gen B}$. 
In the fundamental representation the terms $\mathcal{T}_{\gen{H}\hat{\gen{B}}}+\mathcal{T}_{\gen{M}\gen{b}}$ mix, but we can distinguish them by studying the coproduct both in the $\varrho_\smL\otimes\varrho_\smL$ and in the $\varrho_\smL\otimes\varrho_\smR$ fundamental representations.
One may check that the above coproduct is a homomorphism for the commutation relations with $\gen J$ in both such representations.

\vspace{12pt}

Commutation relations do not fix $\mathcal T_1$, the contribution to the tail which is proportional to the identity operator. At the same time, the freedom of choosing $\mathcal T_1$ may be used to make sure that $\Delta(\gen J)$ is a symmetry of the $R$-matrix. For example, in the $\varrho_\smL\otimes\varrho_\smR$ fundamental representation we can check that\footnote{Here the subscripts LR and RL are used to denote the relevant representations.}
\be
\begin{aligned}
\Delta^{op}_{\smR\smL}(\gen{J})R_{\smL\smR}-R_{\smL\smR}\Delta_{\smL\smR}(\gen{J})&=
i\left[(h_1- {s}_{12})\partial_{p_1}+(h_2+{s}_{12})\partial_{p_2}\right]R_{\smL\smR}
+\mathcal{T}^{op}_{\smR\smL}R_{\smL\smR}-R_{\smL\smR}\mathcal{T}_{\smL\smR}\\
&=(f_{\smL\smR}+\mathcal{T}_{1,\smR\smL}^{op}-\mathcal{T}_{1,\smL\smR})\, R_{\smL\smR}.
\end{aligned}
\ee
Notice the appearance of both $\mathcal{T}_{1,\smL\smR}$ and $\mathcal{T}_{1,\smR\smL}$, because of the opposite coproduct. A similar equation with just the labels L$\leftrightarrow$R swapped is obtained when considering the representation $\varrho_\smR\otimes\varrho_\smL$. If we impose LR symmetry\footnote{Imposing at the same time LR symmetry and (braiding and physical) unitarity singles out a particular class of normalisation for the LR and RL blocks of the $R$-matrix, see~\cite{Borsato:2012ud,Borsato:2013qpa}. A different normalisation of the $R$-matrix results just in a shift of $f_{\smL\smR}$ or $f_{\smR\smL}$ as explained later.} $R_{\smL\smR}=R_{\smR\smL}$ as in~\cite{Borsato:2012ud,Borsato:2013qpa}, we find $f_{\smL\smR}=f_{\smR\smL}$  and we may impose also $\mathcal{T}_{1,\smL\smR}=\mathcal{T}_{1,\smR\smL}$.
The crucial point here is that $f_{\smL\smR}$ is a \emph{scalar} factor. We omit its explicit expression, that is not illuminating nor important for the discussion.
Then boost invariance follows by taking $\mathcal{T}_{1,\smL\smR}=f_{\smL\smR}/2+\mathcal{T}^{symm}_{\smL\smR}$, where $\mathcal{T}^{symm}_{\smL\smR}$ is a contribution symmetric under ``${op}$'' which drops out from the equations.
The computation proceeds similarly for the $\varrho_\smL\otimes\varrho_\smL$  representation, where one finds a corresponding scalar factor $f_{\smL\smL}$.

Obviously, from this point of view the solutions depend on the normalisation  of the $R$-matrix. In fact, if in the above example we had normalised the $R$-matrix with a different scalar factor $R_{\smL\smR}'=e^{\Phi_{12}}R_{\smL\smR}$, then boost invariance would translate to $(f_{\smL\smR}+\mathcal{T}_{1,\smL\smR}^{op}-\mathcal{T}_{1,\smL\smR}+\mathbb{D}\Phi_{12})=0$, where $\mathbb{D}\equiv i(h_1- {s}_{12})\partial_{p_1}+i(h_2+{s}_{12})\partial_{p_2}$. In other words, the solution for $\mathcal T_1$ would be further shifted by $\tfrac{1}{2}\mathbb{D}\Phi_{12}$. This consideration should be taken into account when constructing the physical $S$-matrices for the $AdS_3\times S^3\times T^4$ case\footnote{There is currently no proposal for the physical dressing factors that should solve the crossing equations of~\cite{Borsato:2012ud,Borsato:2015mma} in the $AdS_3\times S^3\times S^3\times S^1$ case.} that include the dressing factors of~\cite{Borsato:2013hoa,Borsato:2016xns}.

\vspace{12pt}

It is natural to expect that there should be a \emph{universal} form of $\mathcal T_1$, which should be valid independently of the representation that we consider. However, this does not mean that the above solutions $\mathcal{T}_{1,\smL\smL},\mathcal{T}_{1,\smL\smR}$ found in the fundamental representation should coincide. In fact, $\mathcal T_1$ may receive contributions both from $\gen H$ and $\gen M$, which could be quite complicated---see e.g. the suggestion~\eqref{eq:cobra-J-univ} towards a universal form of the other terms in the coproduct tail coming from the cobracket.
Since their actions differ on L and R and their contributions mix, expressions in terms of $x^\pm_p$ could look quite different on $\varrho_\smL\otimes\varrho_\smL$ and $\varrho_\smL\otimes\varrho_\smR$. A possibility would be to inspect and compare the solutions for $\mathcal T_1$ in the $\varrho_\smL\otimes\varrho_\smL$ and in the $\varrho_\smL\otimes\varrho_\smR$ fundamental representations, when normalising the $R$-matrix with the physical dressing factors, to see if the results suggest a universal form that evaluates as desired on both cases. We plan to return to this issue in the future.

\subsection{Antipode}
In this section we wish to determine the antipode of the boost $\gen J$.
For all other generators $\gen q$, the antipode is implemented\footnote{In (B.11) of~\cite{Borsato:2013qpa} the antipode is implemented differently on supercharges, see also (B.13). In that paper, one only looks at one 2-dimensional representation (i.e. either L or R), and the swapping of L and R is therefore implemented on the labels of the supercharges rather than on the representations. Here we prefer to write the antipode formula in a more standard way. We still agree with (B.10) of~\cite{Borsato:2013qpa}.} by $S(\gen{q}(p))=\mathscr C\, \gen{q}^{st}(\bar p)\, \mathscr C^{-1}$, where $\mathscr C$ is the charge conjugation matrix, $st$ denotes supertransposition and $\bar p$ is the analytic continuation of the momentum to the crossed region.
In the representation $\varrho_\smL\oplus\varrho_\smR=\text{span}\{\phi^\smL,\psi^\smL,\phi^\smR,\psi^\smR\}$ we may choose
\be
\mathscr C=\sigma_1 \otimes \left(
\begin{array}{cc}
 1 & 0 \\
 0 & i \\
\end{array}
\right),
\ee
which shows that charge conjugation is swapping the L and R representations. When crossing, we send $x^\pm\to1/x^\pm$, with the caveat of dealing with more care with the analytic continuation $\gamma_p\to-i(x^+_p)^{-1}(x^+_p/x^-_p)^{1/2}\gamma_p$.
Essentially, under crossing the coefficients $a_p, b_p$ entering the definitions of the supercharges~\eqref{eq:repr-Q} transform as $a_p\to ib_p$ and $b_p\to ia_p$.
With these prescriptions one finds that the antipode acts as $S(\gen q)=-\gen q$ on all supercharges $\gen Q_\smI,\overline{\gen Q}_\smI$, as well as generators $\gen M, \gen H, \gen b$ and $\hat{\gen B}$.

In order to find out how the antipode acts on $\gen J$ we follow the strategy of~\cite{Borsato:2017lpf} and impose\footnote{This follows by one of the axioms of Hopf algebras $\mu \circ(S\otimes \text{id})\circ\Delta=\mathbf{1}\circ \epsilon$ after setting $\epsilon(\gen J)=0$.} 
\be\label{eq:mu-s-delta}
\mu \circ(S\otimes \text{id})\circ\Delta(\gen{J})=0.
\ee
Let us separate the various contributions arising from the different terms that appear in the boost coproduct. The contribution related to $\Delta'(\gen J)$ obviously does not differ from the $AdS_5/CFT_4$ case~\cite{Borsato:2017lpf}
\be\label{eq:muSDelta-prime}
\mu \circ(S\otimes \text{id})\circ\Delta'(\gen{J})=\left(1+\frac{\ell_p}{h_p}\right)(S(\gen{J})+\gen{J}),
\qquad
\ell_p=\frac{g}{2}\, w^2_p\left(\frac{dw_p}{dp}\right)^{-1}(\cos p-1).
\ee
The tail of the boost coproduct contains factors of $(w_1-w_2)^{-1}$ which potentially generate divergences when acting with the multiplication $\mu$. As in $AdS_5/CFT_4$ we therefore need to carefully check that the divergences cancel in order to get a meaningful result. It is interesting to note that the piece of the tail $\mathcal T_{\gen M\gen b}$---which has no counterpart in $AdS_5/CFT_4$---is an essential ingredient in the case of $AdS_3/CFT_2$, since without it the divergences would not cancel.
When applying the multiplication $\mu$ we identify the two spaces appearing in the tensor product---where we have placed representations with same masses $m$---and we take a limit $p_2\to p_1$.
We find
\be
\begin{aligned}
\mu \circ(S\otimes \text{id})\circ \mathcal{T}_{\gen{H}\hat{\gen{B}}}&=
\lim_{p_2\to p_1} \,\frac{1}{w_1-w_2}\left(-h_1 \hat{b}_1 (1+\tan^2\frac{p_1}{2})\right)(\sigma_3\oplus\sigma_3),\\
\mu \circ(S\otimes \text{id})\circ \mathcal{T}_{\gen{M}\hat{\gen{b}}}&=
\lim_{p_2\to p_1} \,\frac{1}{w_1-w_2}\left(-\frac{m\, w_1}{2}\right)(\sigma_3\oplus\sigma_3),\\
\mu \circ(S\otimes \text{id})\circ (\mathcal{T}_\smL+\mathcal{T}_\smR)&=
\lim_{p_2\to p_1} \,\frac{-w_1}{w_1-w_2}
\left(\gen{Q}_\smL \overline{\gen{Q}}_\smL-\overline{\gen{Q}}_\smL \gen{Q}_\smL
+\gen{Q}_\smR \overline{\gen{Q}}_\smR-\overline{\gen{Q}}_\smR \gen{Q}_\smR
\right)+\text{finite}\\
&=\lim_{p_2\to p_1} \,\frac{1}{w_1-w_2}\, m\, w_1(\sigma_3\oplus\sigma_3)+\text{finite}.
\end{aligned}
\ee
Here we wrote the secret symmetry as $\hat{\gen{B}}=\hat{b}_p (\sigma_3\oplus\sigma_3)$. Since $w_p=\frac{2}{m}h_p\hat{b}_p(1+\tan^2\frac{p}{2})$, we find that all divergent terms cancel each other.

The piece of the tail containing the supercharges produces an additional finite contribution arising from the multiplication of factors of $\gen K$, which generate a factor of $(p_1-p_2)$ cancelling the pole.
If we regularise $p_2=p_1+\epsilon$ and then take the limit $\epsilon\to 0$ we find that the finite contribution produced by $\mathcal T_\smL+\mathcal T_\smR$ is
\be
\lim_{\epsilon\to 0}\frac{i\epsilon}{4}\left(1-\frac{w(p_1+\epsilon)}{w(p_1)}\right)^{-1}\left(\{\gen Q_\smL,\overline{\gen Q}_\smL\}+\{\gen Q_\smR,\overline{\gen Q}_\smR\}\right)=d_p \gen 1,
\qquad
d_p\equiv -\frac{i}{4}w_p\left(\frac{dw_p}{dp}\right)^{-1}h_p.
\ee
Now that we have identified all the terms in the equation~\eqref{eq:mu-s-delta} we can solve it to determine the antipode of $\gen J$
\be
S(\gen{J})=-\gen{J}-\left(1+\frac{\ell_p}{h_p}\right)^{-1}\left(c_p+d_p\right)\gen{1}.
\ee
The expression agrees with the one of $AdS_5/CFT_4$, except for a relative factor of 2 in the definition of $d_p$. We have included also a possible finite contribution $c_p$ arising from the central part $\mathcal T_1$ of the tail of the boost coproduct.

Similarly to the discussion in~\cite{Borsato:2017lpf}, we remark that although we have solved~\eqref{eq:mu-s-delta}, the equation where the antipode acts on the second space $\mu \circ(\text{id}\otimes S)\circ\Delta(\gen{J})=0$ should hold as well.
Following calculations similar to the above ones, in that case one would find $S(\gen{J})=-\gen{J}-(1+\ell_p/h_p)^{-1}(c'_p-d_p)\gen{1}$, where $c'_p$ is the contribution from $\mathcal T_1$ possibly different from the previous $c_p$. Notice the change of sign in front of $d_p$. We conclude that we may have a consistent antipode on $\gen J$ only if the contribution of $\mathcal T_1$ is such that the two results agree. An analoguous question was encountered in \cite{Borsato:2017lpf}, and originally left unanswered. It has subsequently become clear that it is always possible to reverse-engineer the tail of the boost coproduct to incorporate the contribution from a dressing phase which is a solution of the crossing equation\footnote{Cf. \cite{Borsato:2017lpf}, revision to appear.}. The same argument applies in this context, which confirms that the boost, although not capable of constraining the dressing factor, is nevertheless a genuine symmetry of the complete S-matrix\footnote{Access to a universal formulation of the boost coproduct would of course allow a first-principle derivation of the dressing phase, however this is not yet available, and a subject for future study.}.  

It would be interesting to see whether it is possible to find such a $\mathcal T_1$, which at the same time makes sure that the boost coproduct is a symmetry of the $R$-matrix normalised with the physical dressing factors of~\cite{Borsato:2013hoa,Borsato:2016xns}.

\section{Semiclassical limit}\label{sec:sem-lim}
We achieve the semiclassical limit by rescaling the generators $\gen{J}\to g\ \gen{J}$ and $\gen{P}\to \gen{P}/g$ and then taking $g\to\infty$. This corresponds to the BMN limit of~\cite{Berenstein:2002jq}, although from our point of view this is really a contraction of the algebra and not just of the representation. We obtain
\be
\begin{aligned}
&\{\gen{Q}_\smL, \overline{\gen{Q}}_\smL\}=\tfrac{1}{2}(\gen{H}+\gen{M}),\qquad\qquad
&& \{\gen{Q}_\smL,\gen{Q}_\smR\}=-\tfrac{1}{2}\gen P,\\
&\{\gen{Q}_\smR, \overline{\gen{Q}}_\smR\}=\tfrac{1}{2}(\gen{H}-\gen{M}),
&& \{\overline{\gen{Q}}_\smL,\overline{\gen{Q}}_\smR\}=-\tfrac{1}{2}\gen P,\\
& [\gen{J},\gen{H}]=i\gen P,
&& [\gen{J},\gen{Q}_\smI]=-\tfrac{i}{2}\  \overline{\gen{Q}}_{\bar \smI},\\
& [\gen{J},\gen P]=i\gen{H},
&&[\gen{J}, \overline{\gen{Q}}_\smI]=-\tfrac{i}{2}\ \gen{Q}_{\bar \smI},
\end{aligned}
\ee
and 
\be\label{eq:sem-comm-b}
\begin{aligned}
&[\hat{\gen B},\gen Q_\smI]= -\hat{\gen Q}_{\smI}-2 \overline{\gen{Q}}_{\bar{\smI}},
\qquad\qquad
&&[\hat{\gen B},\overline{\gen Q}_\smI]= \hat{\overline{\gen Q}}_{\smI}+2 \gen{Q}_{\bar{\smI}}.
\\
&[\gen{b},\gen{Q}_\smL]=-2\gen{Q}_\smL,\quad
&&[\gen{b},\overline{\gen Q}_\smL]=+2\overline{\gen Q}_\smL,\\
&[\gen{b},\gen{Q}_\smR]=+2\gen{Q}_\smR,\quad
&&[\gen{b},\overline{\gen Q}_\smR]=-2\overline{\gen Q}_\smR,
\end{aligned}
\ee
which shows that the deformed algebra turns into a standard classical superalgebra.
{It contains in particular the Poincar\'e algebra in 2 dimensions (spanned by $\gen P,\gen H,\gen J$) as a subalgebra. There is a clear interpretation of the above limit at the level of the worldsheet. In fact, in the strict semiclassical limit only the quadratic part of the Hamiltonian in light-cone gauge survives (see e.g.~\cite{Arutyunov:2009ga}), and the boost invariance on the worldsheet, which was broken by the gauge in the full Hamiltonian, is restored. One may therefore derive the corresponding Noether charge $\gen{J}=\int d\sigma \left(\sigma\mathscr{H}+\tau \mathscr{P}\right)$, where $\gen{H} = \int d\sigma\mathscr{H},  \gen{P} = -\int d\sigma\mathscr{P}$, and $\sigma,\tau$ parameterise the worldsheet. The canonical quantisation of the usual Poisson brackets will then reproduce the above commutation relations involving the boost. We refer to~\cite{Borsato:2017lpf} for the explicit calculations in the $AdS_5/CFT_4$ case.
Our findings concerning the deformed boost invariance at finite $g$ suggest that the symmetry associated to $\gen J$ should be implemented non-locally on the worldsheet, as indicated by the form of the coproduct.}

\vspace{12pt}

The centrally extended $\su(1|1)_\smL\oplus \su(1|1)_\smR$ superalgebra in the semiclassical limit can be obtained as a contraction of $\alg{sl}(1|2)$. The superalgebra $\alg{sl}(1|2)$ is generated by $3\times 3$ matrices $M_{ij}$ with zeros everywhere except 1 at entry $ij$ that are supertraceless  Str$(A)=A_{11}-A_{22}-A_{33}=0$. 
A Serre-Chevalley basis for $\alg{sl}(1|2)$ with both simple roots fermionic may be given by
\be
\begin{aligned}
&\gen e_1={\phantom{-}}M_{21},\qquad
&&\gen f_1=M_{12},\qquad
&&& \gen h_1={\phantom{-}}M_{11}+M_{22},\\
&\gen e_2=-M_{13},\qquad
&&\gen f_2=M_{31},
&&&\gen h_2=-M_{11}-M_{33},
\end{aligned}
\ee
so that
\be\label{eq:comm-rel-sc}
[\gen h_i,\gen h_j]=0,\qquad
[\gen h_i,\gen e_j]=a_{ij}\gen e_j,\qquad
[\gen h_i,\gen f_j]=-a_{ij}\gen f_j,\qquad
\{\gen e_i,\gen f_j\}=\delta_{ij}\gen h_i,
\ee
with a symmetric Cartan matrix $a_{ij}=(\sigma_1)_{ij}$. The two remaining generators may be found by taking $\gen e_{12}=\{\gen e_1,\gen e_2\},\ \gen f_{12}=-\{\gen f_2,\gen f_1\}$. If we identify the above generators with
\be
\begin{aligned}
&\gen{Q}_\smL={\phantom{-}}\sqrt{\frac{\varepsilon}{2}}(\gen f_1+i\gen e_2),\qquad\qquad
&&\gen{Q}_\smR={\phantom{-}}\sqrt{\frac{\varepsilon}{2}}(i\gen e_1+\gen f_2),\\
&\overline{\gen Q}_\smL=-\sqrt{\frac{\varepsilon}{2}}(\gen e_1+i\gen f_2),
&&\overline{\gen Q}_\smR=-\sqrt{\frac{\varepsilon}{2}}(i\gen f_1+\gen e_2),\\
&\gen{H}=i\varepsilon(-\gen e_{12}+\gen f_{12}),
&&\gen P=-i\varepsilon(\gen h_1+\gen h_2),\\
&\gen{J}=-\frac{i}{2}(\gen e_{12}+\gen f_{12}),
&&\gen{M}=-\varepsilon(\gen h_1-\gen h_2).
\end{aligned}
\ee
and then take $\varepsilon\to 0$ we indeed reproduce the (anti)commutation relations of the $q$-Poincar\'e superalgebra in the semiclassical limit. Notice that we have been careful to identify $\gen P$ with a Cartan generator.

One may be tempted to construct $U_q(\alg{sl}(1|2))$ and try to recover the $q$-Poincar\'e superalgebra under study as a contraction of $U_q(\alg{sl}(1|2))$; in other words the idea would be that of closing the following diagram:
\begin{center}
\begin{tikzpicture}[%
    box/.style={outer sep=1pt},
    Q node/.style={inner sep=5pt,outer sep=0pt},
    arrow/.style={-latex}
    ]%
    \node [box] (qPs) at (7cm  , 0cm) {${q\text{-Poincar\'e superalgebra}}$};
    \node [box] (Ps) at (7cm, -2.5cm) {Poincar\'e superalgebra};
    \node [box] (qsl21) at ( 0cm  , 0cm) {$U_q(\alg{sl}(1|2))$};
    \node [box] (s21) at (0cm, -2.5cm) {$\alg{sl}(1|2)$};
    \newcommand{\horshift}{0.09cm,0cm}
    \newcommand{\vershift}{0cm,0.25cm}
    \draw [arrow] ($(qPs)-(\vershift)$) -- ($(Ps)+(\vershift)$) node [pos=0.5,anchor=west,Q node] {$g\to \infty$};
    \draw [arrow] ($(qsl21)-(\vershift)$) -- ($(s21)+(\vershift)$) node [pos=0.5,anchor=east,Q node] {$q\to 1$};
   \draw [arrow] ($(s21)+(0.8cm,0cm)$) -- ($(Ps)-(2cm,0cm)$) node [pos=0.5,anchor=north,Q node] {$\varepsilon\to 0$};
   \draw [arrow,dashed] ($(qsl21)+(1.2cm,0cm)$) -- ($(qPs)-(2.2cm,0cm)$) node [pos=0.5,anchor=north,Q node] {?};
  \end{tikzpicture}
\end{center}
In~\cite{Borsato:2017lpf} it was shown that in the case of $AdS_5/CFT_4$---in that case $\alg{sl}(1|2)$ is replaced by the $\alg{d}(2,1;\alpha)$ superalgebra---the naive limits fail to achieve the desired contraction and to close the diagram corresponding to the one above. Here we are faced with the same mechanism.
The problem lies in the fact that in the $q$-deformed case the (exponentials of the) Cartan elements will appear as 
\be
\varepsilon\ \frac{q^{\gen h_1\pm \gen h_2}-q^{-(\gen h_1\pm \gen h_2)}}{q-q^{-1}},
\ee
where the explicit $\varepsilon$ comes from the normalisation of the generators.
When considering the combination $\gen h_1+\gen h_2$ it appears natural to take $q=e^{w\varepsilon/2}$, so that factors of $e^{i\gen P}$ will naturally appear after taking the $\varepsilon\to 0$ limit. However, this would at the same time leave unwanted factors of $e^{\gen M}$ coming from $\gen h_1-\gen h_2$, which would prevent us to match with the desired superalgebra. 

We should note, however, that the current situation is much simpler than the $AdS_5/CFT_4$ case. There, in fact, the unwanted factors are exponentials of the Cartans of the $\su(2)$ subalgebra, meaning that it is not obvious how to implement the semiclassical limit only at the level of these generators without spoiling other commutation relations. Here, instead, $\gen M$ is a central element of the superalgebra (after taking $\varepsilon\to 0$), in other words it appears only on the right-hand side of anticommutation relations. Therefore, it would be enough to define a new generator $\gen{M}'\equiv  \frac{1}{w}(e^{\frac{w}{2}\gen{M}}-e^{-\frac{w}{2}\gen{M}})$ to mimic the wanted (anti)commutation relations, where $\gen M$ is replaced by $\gen M'$. Although this trick seems to work at the level of commutation relations, we do not expect that it will go through when including also the coproducts.

\section{Universal $r$-matrix}\label{sec:univ-r}
In this section we wish to construct a universal classical $r$-matrix for $AdS_3/CFT_2$. Besides its intrinsic importance, it will also be a necessary tool for the next section, where we will use it to compute the cobracket of the various generators, in particular the boost.

\subsection{Universal $r$-matrix and CYBE}
We want the $r$-matrix to agree with the semiclassical expansion of the quantum $R$-matrix given in~\eqref{eq:R-LL} and~\eqref{eq:R-LR}, i.e. in the $g\to \infty$ limit we should have
\be
R=1+g^{-1}(r+r_0)+\mathcal{O}(g^{-2}),\qquad
r_0=\phi_0 \mathbf{1}\otimes \mathbf{1}.
\ee
The part proportional to the identity, $r_0$, is sensitive to the normalisation and we will not consider it.
To take the semiclassical limit in the fundamental representation we rewrite\footnote{The semiclassical limit for massless representations should be taken with some care. See the end of this section for a discussion on this.}
\be\label{eq:par-xpm-x}
x^\pm =x \left(\sqrt{1-\frac{m^2 x^2}{g^2
   \left(x^2-1\right)^2}}\pm\frac{im\  x}{g(1- x^2)}\right),
\ee
and send $g\to\infty$. After rewriting the semiclassical expansion of the quantum $R$-matrix in terms of the semiclassical spectral parameter $u$ (related to $x$ as $u=x+1/x,\ x=\frac{1}{2}\left(u+\sqrt{u^2-4}\right)$) we find that it  can be written as
\be\label{eq:ads3-cl-r-ev}
\begin{aligned}
r&=\frac{-i}{u_1-u_2}\Big[2 \sum_{\smI=\smL,\smR}(\gen{Q}_\smI\otimes \overline{\gen{Q}}_\smI
-\overline{\gen{Q}}_\smI\otimes \gen{Q}_\smI)
+\frac{u_2}{u_1} \gen{H}\otimes\gen{B}_0+\frac{u_1}{u_2}\gen{B}_0\otimes \gen{H}+\tfrac{1}{2}\left(\gen{M}\otimes \gen{b}+\gen{b}\otimes \gen{M}\right) 
\Big].
\end{aligned}
\ee
All the generators appearing above are assumed to be written in the semiclassical limit. Moreover, $\gen{B}_0$ corresponds to the level 0 of the secret symmetry, so that $\gen{B}_0\sim u^{-1}\hat{\gen B}$.
Crucially, the above expression matches with the semiclassical expansion of $R$ both in the $\varrho_\smL\otimes\varrho_\smL$ and in the $\varrho_\smL\otimes\varrho_\smR$ representations.\footnote{In fact, the terms $\gen{B}_0\otimes \gen{H}$ and $\gen{b}\otimes \gen{M}$ mix, but they can be distinguished by comparing the expansion of the $R$-matrix both in the $\varrho_\smL\otimes\varrho_\smL$ and in the $\varrho_\smL\otimes\varrho_\smR$ representations.}

We will interpret the above result as the $r$-matrix in \emph{evaluation representation}. If we assume that it comes from a universal expression after identifying the charges at each level $n$ as $\gen q_n=u^n \gen q$, it is easy to reverse-engineer a candidate form for the \emph{universal $r$-matrix}
\be\label{eq:ads3-cl-r}
\begin{aligned}
r&=-i\left(2\, r_{\smL}+2\, r_{\smR}+\, r_{\gen{H}\gen{B}}+\tfrac{1}{2}\, r_{\gen{M}\gen{b}}\right),\\
r_{\smI}&=\sum_{n=0}^\infty(\gen{Q}_{\smI,-1-n}\otimes \overline{\gen{Q}}_{\smI,n}
-\overline{\gen{Q}}_{\smI,-1-n}\otimes \gen{Q}_{\smI,n}),\\
 r_{\gen{H}\gen{B}}&= \sum_{n=-1}^\infty\gen{B}_{-1-n}\otimes \gen{H}_n+\sum_{n=1}^\infty\gen{H}_{-1-n}\otimes\gen{B}_{n},\\
 r_{\gen{M}\gen{b}}&=\sum_{n=0}^\infty(\gen{M}_{-1-n}\otimes \gen{b}_{n}+\gen{b}_{-1-n}\otimes \gen{M}_n).
\end{aligned}
\ee
Although the above proposal for a universal expression matches with the known results, it is important to further test it by checking whether it satisfies the classical Yang-Baxter equation (CYBE) without specifying any representation.
To do that we will follow the strategy used in~\cite{Beisert:2007ty} to check the CYBE for the universal $r$-matrix of $AdS_5/CFT_4$. 
We start by noticing that the above $r$-matrix may be rewritten as
\be
r=r^{\text{can}}+{\sf r},
\qquad\qquad
{\sf r}\equiv-i\left(\gen{B}_{0}\otimes \gen{H}_{-1}-\gen{H}_{-1}\otimes\gen{B}_{0}\right),
\ee
where we interpret $r^{\text{can}}$ as the canonical universal $r$-matrix of the loop algebra $\alg{u}(1|1)_\smL \oplus \alg{u}(1|1)_\smR$.
This superalgebra is spanned by the supercharges $\gen Q_\smI, \overline{\gen Q}_\smI$, I=L,R, the central elements $\gen H,\gen M$ and the inner automorphisms $\gen B_0,\gen b$, which are linear combinations of the inner automorphisms acting separately on the two copies of $\alg{u}(1|1)$.
The universal $r$-matrix of the loop algebra  is built according to the generic construction as~\cite{Etingof:1998,Chari:1994pz}
\be
r^{\text{can}}=-i\ \sum_{n=0}^{\infty}T^A_{-1-n}\otimes T^B_{n}\ g_{AB},
\ee
where $T^A_n$ are the generators at level $n$, and $g_{AB}$ is the (inverse of) an invariant non-degenerate bilinear form.\footnote{In our conventions $\dsbl T^A,T^B\dsbr = f^{AB}_{\ \ \ C}T^C$, where $\dsbl ,\dsbr$ denotes (anti)commutator. We also define the metric as $g^{AB}=g(T^A,T^B)$, so that $g_{AB}$ is the inverse metric. The metric is symmetric in the block of bosonic generators, while it is antisymmetric in the block of fermionic generators. The Killing form of $\alg{u}(1|1)_\smL \oplus \alg{u}(1|1)_\smR$ is degenerate.}
To reproduce our $r$ we take 
\be
g(\gen Q_\smI,\overline{\gen Q}_\smJ)=-\frac{1}{2}\delta_{\smI\smJ},\qquad
g(\gen H,\gen B)=1,\qquad
g(\gen M,\gen b)=2,
\ee
and one may check that the above bilinear form is invariant and non-degenerate on  $\alg{u}(1|1)_\smL \oplus \alg{u}(1|1)_\smR$. 
In what follows we will actually consider a \emph{deformation} of the loop algebra of $\alg{u}(1|1)_\smL \oplus \alg{u}(1|1)_\smR$, as suggested by the strategy of~\cite{Beisert:2007ty}. We write the (anti)commutation relations as
\be
\begin{aligned}
&\{\gen{Q}_{\smL,m}, \overline{\gen{Q}}_{\smL,n}\}=\tfrac{1}{2}(\gen{H}_{m+n}+\gen{M}_{m+n}),\qquad\qquad
&& \{\gen{Q}_{\smL,m},\gen{Q}_{\smR,n}\}=-\beta\gen H_{m+n-1},\\
&\{\gen{Q}_{\smR,m}, \overline{\gen{Q}}_{\smR,n}\}=\tfrac{1}{2}(\gen{H}_{m+n}-\gen{M}_{m+n}),
&& \{\overline{\gen{Q}}_{\smL,m},\overline{\gen{Q}}_{\smR,n}\}=-\beta\gen H_{m+n-1},\\
\end{aligned}
\ee
and
\be
\begin{aligned}
&[\gen b_m,\gen{Q}_{\smL,n}]=-2\gen{Q}_{\smL,m+n},\qquad\qquad
&&[\gen b_m, \overline{\gen{Q}}_{\smL,n}]=+2 \overline{\gen{Q}}_{\smL,m+n},\\
&[\gen b_m,\gen{Q}_{\smR,n}]=+2\gen{Q}_{\smR,m+n},\qquad\qquad
&&[\gen b_m, \overline{\gen{Q}}_{\smR,n}]=-2 \overline{\gen{Q}}_{\smR,m+n},\\
&[\gen B_m,\gen{Q}_{\smI,n}]=-\gen{Q}_{\smI,m+n}-2\beta \overline{\gen{Q}}_{\bar{\smI},m+n-1},\qquad\qquad
&&[\gen B_m, \overline{\gen{Q}}_{\smI,n}]=+ \overline{\gen{Q}}_{\smI,m+n}+2\beta \gen{Q}_{\bar{\smI},m+n-1}.
\end{aligned}
\ee
The undeformed loop-algebra is recovered at $\beta=0$. When setting $\beta=1$, instead, we reproduce the superalgebra that is of interest to us; in particular, the commutators involving the secret symmetry $\gen B$ reduce to the ones in~\eqref{eq:sem-comm-b}. To match we also need the identification $\gen H_{-1}\sim \frac{1}{2}\gen P$. 

\vspace{12pt}

We will now prove that $r$ satisfies CYBE at $\beta=1$---we will actually prove it for generic $\beta$.
We use the fact that $r^{\text{can}}$ satisfies CYBE at $\beta=0$; therefore there are only two types of additional contributions to compute:
\begin{enumerate}
\item those proportional to $\beta$ (coming from deformed commutators) when computing 
\be[r^{\text{can}}_{12},r^{\text{can}}_{13}]+[r^{\text{can}}_{13},r^{\text{can}}_{23}]+[r^{\text{can}}_{12},r^{\text{can}}_{23}],\ee
\item
those coming from the ``mixed terms''
\be[{\sf r}_{12},r^{\text{can}}_{13}]+[{\sf r}_{13},r^{\text{can}}_{23}]+[{\sf r}_{12},r^{\text{can}}_{23}]+[r^{\text{can}}_{12},r^{\text{can}}_{13}]+[r^{\text{can}}_{13},{\sf r}_{23}]+[r^{\text{can}}_{12},{\sf r}_{23}].\ee
\end{enumerate}
Notice that terms of the form $[{\sf r},{\sf r}]$ are automatically 0 since $\gen B_m$ and $\gen H_m$ commute.
For contributions of type 1 we find
\be
\begin{aligned}
[r^{\text{can}}_{12},r^{\text{can}}_{13}]:\ &
-4\beta\sum_{m,n=0}^{\infty}
\mathcal X_{[-3-n-m,\ n,\ m]},\\
[r^{\text{can}}_{13},r^{\text{can}}_{23}]:\ &
-4\beta\sum_{m,n=0}^{\infty}
\mathcal X_{[-1-n,\ -1-m,\ m+n-1]},\\
[r^{\text{can}}_{12},r^{\text{can}}_{23}]:\ &
+4\beta\sum_{m,n=0}^{\infty}
\mathcal X_{[-1-n,\ n-m-2,\ m]},\\
\end{aligned}
\ee
where we defined
\be
\begin{aligned}
\mathcal X_{[n_1,n_2,n_3]}\equiv&
\Big(\gen H_{n_1}\otimes \gen Q_{\smL,n_2}\otimes \gen Q_{\smR,n_3}
+\gen Q_{\smL,n_1}\otimes \gen Q_{\smR,n_2}\otimes \gen H_{n_3}\\
&\qquad
-\gen Q_{\smL,n_1}\otimes \gen H_{n_2}\otimes \gen Q_{\smR,n_3}
+\smL \leftrightarrow \smR
\Big)+\gen Q \leftrightarrow \overline{\gen Q}.
\end{aligned}
\ee
To avoid long expressions, here we are not writing explicitly all the terms. For each term that we write explicitly there are three additional ones, obtained by first exchanging labels L$\leftrightarrow$R, and then $\gen Q \leftrightarrow \overline{\gen Q}$ everywhere.
Summing up the above results we obtain
\be
-4\beta\left(\sum_{m=0}^{\infty}\sum_{n=m+2}^{\infty}+\sum_{n=0}^{\infty}\sum_{m=n-1}^{\infty}-\sum_{n=0}^{\infty}\sum_{m=0}^{\infty}\right)\mathcal X_{[-1-n,n-m-2,m]}
=-4\beta\mathcal X_{[-1,-1,-1]},
\ee
where we first relabelled the summed indices, and then used the identity
\be
\left(\sum_{m=0}^{\infty}\sum_{n=m+1}^{\infty}+\sum_{n=0}^{\infty}\sum_{m=n}^{\infty}-\sum_{n=0}^{\infty}\sum_{m=0}^{\infty}\right)F_{mn}=0,
\ee
which is valid due to cancellation of the domains for any collection of objects $F_{mn}$ labelled by $m$ and $n$.
We will now show that $-4\beta\mathcal X_{[-1,-1,-1]}$ is exactly cancelled by the contributions of type 2.
We find
\be
\begin{aligned}
[{\sf r}_{12},r^{\text{can}}_{13}]:\ &
+\sum_{n=0}^\infty \left(
2\gen Q_{\smL,-1-n}\otimes \gen H_{-1}\otimes \overline{\gen Q}_{\smL,n}
+4\beta \gen Q_{\smL,-2-n}\otimes \gen H_{-1}\otimes \gen Q_{\smR,n}
+\smL \leftrightarrow \smR 
\right)+\gen Q \leftrightarrow \overline{\gen Q},\\
[r^{\text{can}}_{12},{\sf r}_{13}]:\ &
-\sum_{n=0}^\infty \left(
2\gen Q_{\smL,-1-n}\otimes \overline{\gen Q}_{\smL,n}\otimes \gen H_{-1}
+4\beta \gen Q_{\smL,-2-n}\otimes \gen Q_{\smR,n}\otimes \gen H_{-1}
+\smL \leftrightarrow \smR 
\right)+\gen Q \leftrightarrow \overline{\gen Q},\\
[{\sf r}_{13},r^{\text{can}}_{23}]:\ &
+\sum_{n=0}^\infty \left(
2\gen H_{-1}\otimes \gen Q_{\smL,-1-n}\otimes \overline{\gen Q}_{\smL,n}
+4\beta  \gen H_{-1}\otimes\gen Q_{\smL,-1-n}\otimes \gen Q_{\smR,n-1}
+\smL \leftrightarrow \smR 
\right)+\gen Q \leftrightarrow \overline{\gen Q},\\
[r^{\text{can}}_{13},{\sf r}_{23}]:\ &
-\sum_{n=0}^\infty \left(
2\gen Q_{\smL,-1-n}\otimes \gen H_{-1}\otimes \overline{\gen Q}_{\smL,n}
+4\beta  \gen Q_{\smL,-1-n}\otimes \gen H_{-1}\otimes\gen Q_{\smR,n-1}
+\smL \leftrightarrow \smR 
\right)+\gen Q \leftrightarrow \overline{\gen Q},\\
[{\sf r}_{12},r^{\text{can}}_{23}]:\ &
-\sum_{n=0}^\infty \left(
2\gen H_{-1}\otimes \gen Q_{\smL,-1-n}\otimes \overline{\gen Q}_{\smL,n}
+4\beta  \gen H_{-1}\otimes\gen Q_{\smL,-2-n}\otimes \gen Q_{\smR,n}
+\smL \leftrightarrow \smR 
\right)+\gen Q \leftrightarrow \overline{\gen Q},\\
[r^{\text{can}}_{12},{\sf r}_{23}]:\ &
+\sum_{n=0}^\infty \left(
2\gen Q_{\smL,-1-n}\otimes \overline{\gen Q}_{\smL,n}\otimes \gen H_{-1}
+4\beta \gen Q_{\smL,-1-n}\otimes \gen Q_{\smR,n-1}\otimes \gen H_{-1}
+\smL \leftrightarrow \smR 
\right)+\gen Q \leftrightarrow \overline{\gen Q}.\\
\end{aligned}
\ee
It is easy to see that all $\beta$-independent terms  cancel each other, while the $\beta$-dependent ones leave a finite result due to some shifts in the levels in some expressions. The result
\be
4\beta \sum_{n=0}^\infty \left(\mathcal X_{[-1,-1-n,n-1]}-\mathcal X_{[-1,-2-n,n]}\right)=4\beta \mathcal X_{[-1,-1,-1]}
\ee
exactly cancels the contributions of type 1, and the CYBE is checked for generic $\beta$. Notice that, for the calculation to work, it was crucial to have level shifts of one unity in the $\beta$-dependent terms of the (anti)commutation relations, as well as the additional ${\sf r}$.

\subsection{Massless representations, semiclassical limit and the $r$-matrix}
The parameterisation~\eqref{eq:par-xpm-x} of the Zhukovski variables is not adequate in the massless limit $m\to 0$, since it would imply $x^+=x^-$ and $p=2\pi n$. A different parameterisation is therefore needed in the massless case, and we can find it e.g. by sending $m\to 0$ only after redefining $x=1+\frac{m}{2\xi}$ (or $x=-1-\frac{m}{2\xi}$) in~\eqref{eq:par-xpm-x}.
We find
\be\label{eq:par-xpm-x-0}
x^\pm = \pm \frac{i\xi}{g}+\sqrt{1-\frac{\xi^2}{g^2}},
\qquad\text{  or  }\qquad
x^\pm = \pm \frac{i\xi}{g}-\sqrt{1-\frac{\xi^2}{g^2}},
\ee
where the first parameterisation implies\footnote{Here we assume $-\pi<p<\pi$} $p>0$ while the second one $p<0$. Therefore, we need to distinguish between \emph{worldsheet left- and right-movers}.
In both cases the energy is $2g\, \sin( p/2)=2\xi$. The coefficients parameterising the supercharges in~\eqref{eq:repr-Q} are just $a_p=-b_p=\sqrt{\xi}$ in the first parameterisation, and $a_p=+b_p=\sqrt{\xi}$ in the second one. Let us emphasise that we have not taken the $g\to \infty$ limit yet.
Notice that the secret symmetry in~\eqref{eq:secret} vanishes in the massless limit, since $x^+=1/x^-$ when $m=0$.
Furthermore, the spectral parameter $\hat u=(x^++x^-+1/x^++1/x^-)/2$ reduces to  $\pm 2\sqrt{1-\xi^2/g^2}$, where the sign $\pm$ depends on which of the above parameterisations is chosen.
Therefore, semiclassically $\hat u\to u=\pm 2$.

Let us make a comment on the $g$-dependence. In the massless case we may parameterise $x^\pm =e^{\pm ip/2}$, so that there is no explicit $g$-dependence. This is not a good parameterisation if we want to take a semiclassical limit $g\to \infty$, since for example the massless--massless $R$-matrix would not expand as $1+\mathcal{O}(1/g)$. If instead we use the parameterisation above in terms of $\xi$, we reintroduce the missing $g$-dependence, and it makes sense to expand our results at large $g$. This is similar to what one does in the BMN limit~\cite{Berenstein:2002jq}, where one first rescales $p\to p/g$.

\vspace{12pt}

Let us now discuss the semiclassical limit of the $R$-matrix. First we consider the case of massless--massive scattering, where the mass of the second excitation is generic but not 0. To obtain the classical $r$-matrix in the fundamental representation  we first consider the massless--massive $R$-matrix, where $x^\pm_1$ are parameterised in terms of~\eqref{eq:par-xpm-x-0} and  $x^\pm_2$ in terms of~\eqref{eq:par-xpm-x}. Then we send $g\to \infty$ and we obtain $R=1+r/g+\mathcal{O}(1/g^2)$.
We have checked that what we obtain coincides with the $r$-matrix in evaluation representation as written in~\eqref{eq:ads3-cl-r-ev}.

Particular care is needed when taking the semiclassical limit in the case of massless--massless scattering. In fact, we must scatter a left- with a right-mover, i.e. we must use the first parameterisation in~\eqref{eq:par-xpm-x-0} for one excitation and the second one for the other. From the operational point of view, this is done to avoid the appearance of infinities. Physically it is justified by the fact that we want the two massless excitations to travel in opposite directions, so that they have the chance to meet, since they both go at the speed of light.
Then we extract the classical $r$-matrix from the semiclassical expansion of the massless--massless $R$-matrix,  $R=1+r/g+\mathcal{O}(1/g^2)$.
We get $r=-i\sqrt{\xi_1}\sqrt{\xi_2}M$, where $M=\sigma_+\otimes\sigma_-+\sigma_-\otimes\sigma_+$.
Also this result matches with the classical $r$-matrix written in evaluation representation in~\eqref{eq:ads3-cl-r-ev}, and this can be seen quite simply. In fact, all terms in $r$ containing $\gen M$ or $\gen B$ obviously vanish. The only contributions come from the supercharges, and
\be
\gen{Q}_\smL\otimes \overline{\gen{Q}}_\smL-\overline{\gen{Q}}_\smL\otimes \gen{Q}_\smL
+\gen{Q}_\smR\otimes \overline{\gen{Q}}_\smR-\overline{\gen{Q}}_\smR\otimes \gen{Q}_\smR=
(a_1\bar a_2-\bar b_1 b_2)\sigma_-\otimes \sigma_+- (\bar a_1 a_2-b_1 \bar b_2)\sigma_+\otimes \sigma_-,
\ee
where we have used the parameterisation coefficients as in~\eqref{eq:repr-Q}.
Now it is crucial that we are taking two massless excitations in opposite kinematical regimes, i.e. $a_1=-b_1$ and $a_2=+b_2$. Recalling that in our parameterisation we  have \emph{real} coefficients ($\bar a=a, \bar b=b$), this means that the contributions add up instead of cancelling $(a_1\bar a_2-\bar b_1 b_2)= 2a_1a_2=(\bar a_1 a_2-b_1 \bar b_2)$.  Using $u_1=-u_2=2$ we obtain $r=-i\sqrt{\xi_1}\sqrt{\xi_2}M$ as wanted.

\section{Cobracket}\label{sec:cobra}
Although most part of the results---e.g. the $R$-matrix in~\eqref{eq:R-LL} and~\eqref{eq:R-LR} and the coproducts in~\eqref{eq:Delta-su(1|1)2} and~\eqref{eq:start-Delta-boost}---are only given in the fundamental representation, the universal $r$-matrix proposed in the previous section allows us to go towards a universal formulation. In particular, if we consider the invariance of the $R$-matrix under a generic generator $\gen q$ as in~\eqref{eq:inv-R} and we implement a semiclassical expansion, we find that $\delta(\gen{q})\equiv\Delta_{(1)}(\gen{q})-\Delta_{(1)}^{op}(\gen{q})$ may be obtained by computing the commutator $\delta(\gen{q})=[\gen{q}\otimes \gen 1+\gen 1\otimes \gen{q},r]$. One gets this result  after expanding the coproduct as $\Delta (\gen q) = \gen{q}\otimes \gen 1+\gen 1\otimes \gen{q}+g^{-1}\Delta_{(1)}(\gen{q})+\mathcal{O}(g^{-2})$.
In other words $\delta(\gen{q})$, which we call the cobracket of $\gen q$, can be derived in universal form thanks to the knowledge of the universal $r$-matrix.

We present the results for the cobrackets of all the generators of the deformed loop algebra $\alg{u}(1|1)_\smL \oplus \alg{u}(1|1)_\smR$ of the previous section.
In universal form they read
\begin{align}
\delta({\bf Q}_{\smL,m}) & = \phantom{-} i \sum_{n=0}^m \left[ {\bf H}_{m-n-1} \otimes {\bf Q}_{\smL,n} - {\bf Q}_{\smL,m-n} \otimes {\bf H}_{n-1} \right] \notag \\
& \phantom{=} \, \, + i \sum_{n=0}^{m-1} \Big[ 2 \beta \left( {\bf H}_{m-n-2} \otimes {\bf \overline{Q}}_{\smR,n} - {\bf \overline{Q}}_{\smR,m-n-1} \otimes {\bf H}_{n-1} \right) \notag \\
& \qquad\qquad\quad + {\bf M}_{m-n-1} \otimes {\bf Q}_{\smL,n} - {\bf Q}_{\smL,m-n-1} \otimes {\bf M}_n \Big], \\
\delta({\bf \overline{Q}}_{\smL,m}) & = - i \sum_{n=0}^m \left[ {\bf H}_{m-n-1} \otimes {\bf \overline{Q}}_{\smL,n} - {\bf \overline{Q}}_{\smL,m-n} \otimes {\bf H}_{n-1} \right]\notag \\
& \phantom{=} \, \,  - i \sum_{n=0}^{m-1} \Big[ 2 \beta \left( {\bf H}_{m-n-2} \otimes {\bf Q}_{\smR,n} - {\bf Q}_{\smR,m-n-1} \otimes {\bf H}_{n-1} \right) \notag \\
& \qquad\qquad\quad+ {\bf M}_{m-n-1} \otimes {\bf \overline{Q}}_{\smL,n} - {\bf \overline{Q}}_{\smL,m-n-1} \otimes {\bf M}_n \Big], 
\end{align}
\begin{align}
\delta({\bf Q}_{\smR,m}) & = \phantom{-} i \sum_{n=0}^m \left[ {\bf H}_{m-n-1} \otimes {\bf Q}_{\smR,n} - {\bf Q}_{\smR,m-n} \otimes {\bf H}_{n-1} \right]  \notag \\
& \phantom{=} \, \, + i \sum_{n=0}^{m-1} \Big[ 2 \beta \left( {\bf H}_{m-n-2} \otimes {\bf \overline{Q}}_{\smL,n} - {\bf \overline{Q}}_{\smL,m-n-1} \otimes {\bf H}_{n-1} \right) \notag \\
& \qquad\qquad\quad- {\bf M}_{m-n-1} \otimes {\bf Q}_{\smR,n} + {\bf Q}_{\smR,m-n-1} \otimes {\bf M}_n \Big], \\
\delta({\bf \overline{Q}}_{\smR,m}) & =  - i \sum_{n=0}^m \left[ {\bf H}_{m-n-1} \otimes {\bf \overline{Q}}_{\smR,n} - {\bf \overline{Q}}_{\smR,m-n} \otimes {\bf H}_{n-1} \right]  \notag \\
& \phantom{=} \, \,- i \sum_{n=0}^{m-1} \Big[ 2 \beta \left( {\bf H}_{m-n-2} \otimes {\bf Q}_{\smL,n} - {\bf Q}_{\smL,m-n-1} \otimes {\bf H}_{n-1} \right) \notag \\
& \qquad\qquad\quad - {\bf M}_{m-n-1} \otimes {\bf \overline{Q}}_{\smR,n} + {\bf \overline{Q}}_{\smR,m-n-1} \otimes {\bf M}_n \Big], 
\end{align}
\begin{align}
\delta({\bf b}_m) & = 4 i \sum_{n=0}^{m-1} \Big[ {\bf Q}_{\smL,m-n-1} \otimes {\bf \overline{Q}}_{\smL,n} - {\bf Q}_{\smR,m-n-1} \otimes {\bf \overline{Q}}_{\smR,n}\notag \\
& \qquad\qquad\quad + {\bf \overline{Q}}_{\smL,m-n-1} \otimes {\bf Q}_{\smL,n} - {\bf \overline{Q}}_{\smR,m-n-1} \otimes {\bf Q}_{\smR,n} \Big] , \\
\delta({\bf B}_m) & = 2i \sum_{\smI=\smL,\smR}\sum_{n=0}^{m-1} \left[ {\bf Q}_{\smI,m-n-1} \otimes {\bf \overline{Q}}_{\smI,n} + {\bf \overline{Q}}_{\smI,m-n-1} \otimes {\bf Q}_{\smI,n} \right] \notag \\
& \phantom{=} + 4 i \beta \sum_{\smI=\smL,\smR}\sum_{n=0}^{m-2} \left[ {\bf Q}_{\bar{\smI},m-n-2} \otimes {\bf Q}_{\smI,n} + {\bf \overline{Q}}_{\bar{\smI},m-n-2} \otimes {\bf \overline{Q}}_{\smI,n} \right] .
\end{align}
Notice that the cobrackets of the barred supercharges are obtained through complex conjugation of their non-barred correspondences, together with the exchange ${\bf Q}_{\smI} \leftrightarrow {\bf \overline{Q}}_{\smI}$. Also, the signs of the terms involving the generator ${\bf M}$ keep us from easily writing the cobrackets of the supercharges in the more compact forms $\delta({\bf Q}_{\smI,m})$ and $\delta({\bf \overline{Q}}_{\smI,m})$.
{Since $\gen M_m,\gen H_m$ are central elements of the loop algebra of $\alg{u}(1|1)_\smL \oplus \alg{u}(1|1)_\smR$, their cobrackets are trivial.}

\subsection{Cobracket of the boost}
We now wish to compute the cobracket of the boost  $\delta(\gen{J})=[\gen{J}\otimes \gen 1+\gen 1\otimes \gen{J},r]$. As in~\cite{Borsato:2017lpf} we use 
\be
[\gen{J},\gen{B}_0]=-2i\, \gen{B}_{-1},
\ee
which is motivated by the fundamental representation. Moreover, if we define the action of the boost on a generic generator $\gen q_0$ at level 0 as $\tilde{\gen{q}}_0\equiv[\gen{J},\gen{q}_0]$, we will assume that the boost acts on the level $n$ $\gen{q}_n\sim u^n \gen q_0$ as
\be\label{eq:Jqn}
[\gen{J},\gen{q}_n]=\tilde{\gen{q}}_n+in\left(2\gen{q}_{n-1}-\tfrac{1}{2}\gen{q}_{n+1}\right),
\ee
where $\tilde{\gen{q}}_n\sim u^n\tilde{\gen{q}}_0$.
This commutator is justified by the result in the evaluation representation.
In universal form we find
\be\label{eq:cobra-J-univ}
\begin{aligned}
\delta(\gen{J})=&-i\left(2\, \delta_{\smL}(\gen{J})+2\, \delta_{\smR}(\gen{J})+\, \delta_{\gen{H}\gen{B}}(\gen{J})+\tfrac{1}{2}\, \delta_{\gen{M}\gen{b}}(\gen{J})\right),\\
\delta_{\smI}(\gen{J})=&
i\sum_{m=0}^\infty \left[ \gen{Q}_{\smI,-m} \otimes \overline{\gen{Q}}_{\smI,m}
- \overline{\gen{Q}}_{\smI,-m}\otimes \gen{Q}_{\smI,m}\right]
-\frac{i}{2}\left[ \gen{Q}_{\smI,0} \otimes \overline{\gen{Q}}_{\smI,0}
- \overline{\gen{Q}}_{\smI,0}\otimes \gen{Q}_{\smI,0}\right],\\
\delta_{\gen{H}\gen{B}}(\gen{J})=&i\Bigg(\sum_{m=0}^\infty \gen{B}_{-m}\otimes \gen{H}_m
+\sum_{m=1}^\infty \gen{H}_{-m}\otimes \gen{B}_m\Bigg),\\
\delta_{\gen{M}\gen{b}}(\gen{J})=&
i\sum_{m=0}^\infty \left[ \gen{M}_{-m} \otimes \gen{b}_{m}
+ \gen{b}_{-m} \otimes \gen{M}_{m}\right]
-\frac{i}{2} \left[ \gen{M}_{0} \otimes \gen{b}_{0}
+ \gen{b}_{0} \otimes \gen{M}_{0}\right].
\end{aligned}
\ee
As expected, $\delta_{\gen{H}\gen{B}}(\gen{J})$ is identical\footnote{In the above expression we have already used the identification $\gen{P}_n \sim 2\gen{H}_{n-1}$. We refer to~\cite{Borsato:2017lpf} for the expression of $\delta_{\gen{H}\gen{B}}(\gen{J})$ before this identification.} to the case of $AdS_5/CFT_4$, and one may notice close similarities also in the contributions with the supercharges.
After going to evaluation representation we obtain
\be
\begin{aligned}
\delta(\gen J)&=
\frac{u_1+u_2}{u_1-u_2}\Bigg(\sum_{\smI=\smL,\smR}(\gen{Q}_\smI\otimes \overline{\gen{Q}}_\smI- \overline{\gen{Q}}_\smI\otimes \gen{Q}_\smI)
+\tfrac{1}{4}\left(\gen{M}\otimes \gen{b}+\gen{b}\otimes \gen{M}\right) 
\Bigg)\\
&\qquad+\frac{1}{u_1-u_2}\left(\gen{H}\otimes \gen{B}_1+\gen{B}_1\otimes \gen{H}\right).
\end{aligned}
\ee
There are obvious analogies between the cobracket and the exact coproduct given in~\eqref{eq:start-Delta-boost},~\eqref{eq:Deltap} and~\eqref{eq:DeltaJ}. As in the case of $AdS_5/CFT_4$, the result suggests that the semiclassical spectral parameter $u_i$ is replaced at the quantum level by $w_i$. Certain terms in the exact coproduct $\Delta(\gen J)$ are not captured by the cobracket, either because they start entering at orders higher than $1/g$ (e.g. the contribution with $\tan\frac{p}{2}\otimes \tan\frac{p}{2}$ in $\mathcal{T}_{\gen{H}\hat{\gen{B}}}$) or because they are symmetric under the action of ``${op}$'' (e.g. the correction to the trivial coproduct in $\Delta'(\gen J)$).

\section{Conclusions}
In this paper we have shown that the $q$-Poincar\'e supersymmetry is not exclusive to the $AdS_5/CFT_4$ integrable problem, and that it can be realised also in $AdS_3/CFT_2$. This suggests that, similarly to what happened for the secret symmetry, also the invariance under the boost $\gen J$ should be viewed as one of the several common features shared by the $AdS/CFT$ integrable models.  It would be interesting to identify other manifestations of $\gen J$ in $AdS/CFT$.
{
In particular, a background recently found to be integrable is $AdS_2 \times S^2 \times T^6$, with superisometry $\alg{psu}(1,1|2)$. The holographic dual might either be a superconformal quantum mechanics, or a chiral CFT \cite{Sorokin:2011rr,Murugan:2012mf}. In \cite{Hoare:2014kma} an exact $S$-matrix theory was built, realising a centrally-extended $\alg{psu}(1|1)$ Lie superalgebra. Yangian, bonus symmetry and Bethe ansatz have been studied in \cite{Hoare:2014kma,Hoare:2015kla,Fontanella:2017rvu,Torrielli:2017nab}.
 On the one hand, observing  the boost symmetry also in the $AdS_2$ case, which appears to be amenable to a similar algebraic treatment as its higher-dimensional analogues, would confirm the universal nature of the symmetry we are finding. On the other hand, the $AdS_2$ integrable structure is in several ways more subtle, therefore progress towards the complete solution of the model is harder to come, and it is decorated with open questions. Discovering the boost symmetry in that setup could represent a crucial step in overcoming some of these open problems, and we plan to return to this issue in future work.
}

Since the action of $\gen J$ includes taking a derivative with respect to the worldsheet momentum, the boost invariance is sensitive to the normalisation of the $S$-matrix. Nevertheless, a different normalisation of $S$ would produce only a shift in the tail of $\Delta(\gen J)$ proportional to the identity matrix. Since we can only reverse-engineer the boost coproduct and we cannot fix it \emph{a priori}, we cannot obtain constraints on the dressing phases of $AdS_3/CFT_2$. In a scenario where the boost coproduct were instead known in universal form, the dressing phases would need to satisfy certain differential equations, and one could further test the proposals of~\cite{Borsato:2013hoa,Borsato:2016xns}. It would be therefore very interesting to find alternative ways to fix the tail of the boost coproduct, including its contribution proportional to the identity. The achievement of this goal would certainly require some additional inputs, and the specification of which $AdS_3$ background is studied. In fact the dressing phases of $AdS_3\times S^3\times S^3\times S^1$ and of $AdS_3\times S^3\times T^4$ are expected to be different.

{Let us mention that the $AdS_3$ backgrounds that we are considering can in general be supported by a mixture of Neveu-Schwarz--Neveu-Schwarz (NSNS) and Ramond--Ramond (RR) fluxes. It is known that in the generic case the off-shell algebra is essentially the same as the one in the pure RR case considered here, and that the representations will depend on an additional parameter $\k$ corresponding to the relative amount of the fluxes~\cite{Lloyd:2014bsa,Borsato:2015mma}.\footnote{It may be useful, especially when attempting to include the boost, to reformulate the construction and have already the commutation relations, rather than just the representations, deformed by this additional parameter. For example, instead of having $\{\gen Q_\smL,\overline{\gen Q}_\smL \} = \tfrac{1}{2}(\gen H+\gen M)$ where $\gen M$ has eigenvalues $m + \k p$~\cite{Lloyd:2014bsa,Borsato:2015mma}, one may prefer to write $\{\gen Q_\smL,\overline{\gen Q}_\smL \} = \tfrac{1}{2}(\gen H+\gen M+\k \gen P)$ where $\gen M$ has eigenvalues $m$. The generator $\gen M$ would then remain central even when including a generator acting as the derivative with respect to momentum.} It would be interesting to extend the deformed boost invariance to the generic case of mixed fluxes: since the dispersion relation depends on $\k$, that would correspond to a deformation of the $q$-Poincar\'e algebra considered here, and it would be nice to investigate it also in the pure NSNS limit.}

Motivated by the desire of better understanding  the boost symmetry, we have also proposed a universal expression for the classical $r$-matrix of the  $AdS_3/CFT_2$ integrable system. Its structure resembles the one of the $AdS_5/CFT_4$ classical $r$-matrix of Beisert and Spill~\cite{Beisert:2007ty}. With this result we complete some information that was missing in $AdS_3/CFT_2$, and we contribute to put $AdS_3/CFT_2$ in a status closer to the one of its higher dimensional cousin.

{
The appearance of the boost invariance in both $AdS_5/CFT_4$ and $AdS_3/CFT_2$  gives us further confidence that $\gen J$ should not be just an accidental symmetry of the fundamental representations of the underlying symmetries, and that $\gen J$ may help to shed some light on the universal formulation of the corresponding quantum groups.
}

\subsection*{Note added}
While writing this manuscript we received the interesting paper~\cite{Pittelli:2017spf}, where the universal classical $r$-matrix of $AdS_3/CFT_2$ is justified from the $R\mathcal{TT}$ formulation. In order to match our results with those of~\cite{Pittelli:2017spf} we need to identify the generators as
\be\label{eq:dictionary-Antonio}
\begin{aligned}
& \mathfrak{Q}_L^{(n)} = - \gen Q_{\smL,n}, && \mathfrak S_L^{(n)} = \overline{\gen{Q}}_{\smL,n}, 
&&& \mathfrak H_L^{(n)} = -\tfrac{1}{2} (\gen H_n+\gen M_n), &&&&  \mathfrak H_R^{(n)} = -\tfrac{1}{2} (\gen H_n-\gen M_n),\\
&\mathfrak{Q}_R^{(n)} = - \gen Q_{\smR,n}, && \mathfrak S_R^{(n)} = \overline{\gen{Q}}_{\smR,n},
&&& \ss_L^{(n)} =  -\tfrac{1}{2}\gen b_{n+1} - \gen B_{n+1}, &&&&  \ss_R^{(n)} =  \tfrac{1}{2}\gen b_{n+1} - \gen B_{n+1},
\end{aligned}
\ee
where the notation of each paper is used, and the identification of $\gen b_n, \gen B_n$ is to be understood up to central elements.

\section{Acknowledgments}

We would like to thank Antonio Pittelli for discussions, and for sharing with us a copy of the manuscript ``Yangian Symmetry of String Theory on $AdS_3 \times S^3 \times S^3 \times S^1$ with Mixed 3-form Flux"~\cite{Pittelli:2017spf} whilst in preparation. We thank Ben Hoare, Marius de Leeuw and Juan Miguel Nieto for discussions. {We also thank Olof Ohlsson Sax and Antonio Pittelli for comments on the manuscript.}
The work of R.B. was supported by the ERC advanced grant No 341222. J.S. and A.T. thank the STFC under the Consolidated Grant project nr. ST/L000490/1. 

\subsection*{Data Management}

No data beyond those presented in this paper are needed to validate its findings.

\bibliographystyle{nb}
\bibliography{biblio}{}

\end{document}